\documentstyle[12pt,epsfig,dcolumn,amssymb]{article}
\def\reference{\parskip 0pt\par\noindent\hangindent 0.5 truecm}
\def\kms{km ${\rm s}^{-1}$}
\def\HI{H{\small I} }
\def\HIn{H{\small I}}%
\newcommand{\expect}[1]{\langle{#1}\rangle}

\newlength{\numberwidth}
\newcommand{\nw}{\hspace*{\numberwidth}}
\newlength{\onewidth}
\newlength{\dotwidth}
\newcommand{\dw}{\hspace*{\dotwidth}}
\newlength{\minuswidth}
\newlength{\asteriskwidth}
\newlength{\comparisonwidth}
\newlength{\pmwidth}
\newcommand{\pw}{\hspace*{\pmwidth}}

\begin{document}

\title{The HI Content of Compact Groups of Galaxies}

\author{J.B.~Stevens$^1$,
  R.L.~Webster$^2$,
  D.G.~Barnes$^2$,
  D.J.~Pisano$^3$, \and
  M.J.~Drinkwater$^4$
}

\date{}
\maketitle

{\center
$^1$School of Physics, University of Melbourne, Victoria, Australia 3010\\Affiliated with the Australia Telescope National Facility, CSIRO\\[3mm]
$^2$School of Physics, University of Melbourne, Victoria, Australia 3010\\[3mm]
$^3$NSF MPS Distinguished Research Fellow,Australia Telescope National Facility, Epping, New South Wales, Australia 1710\\[3mm]
$^4$Department of Physics, University of Queensland, Queensland, Australia 4072\\[3mm]
}

\begin{abstract}
The \HI content of Hickson Compact Groups in the southern hemisphere 
is measured using data from the \HI Parkes All-Sky Survey (HIPASS), and
dedicated observations using the narrow band filter on the Multibeam 
instrument on the Parkes telescope. The expected \HI mass
of these groups was estimated using the luminosity, diameter and
morphological types of the member galaxies, calibrated from published
data. Taking careful account of non-detection limits, the results show that the compact group population that has been
detected by these observations has an \HI content similar to that of galaxies in the
reference field sample. The upper limits for the undetected groups lie
within the normal range; improvement of these limits will require a large 
increase in sensitivity.
\end{abstract}


\bigskip

\section{Introduction}
Neutral hydrogen gas (\HIn) on the edges of spiral galaxies is only very loosely gravitationally bound. Therefore, when spiral galaxies interact, the outer \HI can be easily disturbed (Hibbard \& Van Gorkom 1996). During a merger, the \HI may be heated and ionised, or may cool into molecular (${\rm H}_{2}$) clouds within the remnant (Hibbard \& Van Gorkom 1996). It may be tidally driven into the centre of the galaxy and converted into molecular gas, or it could be turned into stars when tidal streams collide and become compressed. The gas may also escape the interacting system and disperse, becoming too diffuse to detect, or perhaps ionised. This paper re-examines the evidence for \HI deficiency in compact groups of galaxies using a new survey, the \HI Parkes All Sky Survey.

Compact groups of galaxies are excellent places to look for interactions and mergers between (spiral) galaxies. A compact group is defined as having several galaxies with similar redshifts within a small area of the sky, which are also isolated from surrounding galaxies. This ensures that the cores of clusters are excluded from the definition.

A compact group generally has a low velocity dispersion, $\sim$250 km s$^{-1}$ (Hickson 1997), and a crossing time much smaller than the Hubble time, meaning interactions and mergers are very likely. Although the fraction of spirals in the field is higher ($\sim$80\%), the fraction of compact group galaxies which are spirals is still $\sim$50\% (Hickson 1997).

Although interactions are thought to be common in compact groups, the star-formation rate (SFR), and the amount of molecular gas in the galaxies in the compact groups, are only weakly enhanced in comparison to field galaxies (Leon et al. 1998). They concluded that only compact groups with a very small mean separation of $<$30 kpc had a strong ${\rm H}_{2}$ enhancement. Verdes-Montenegro et al. (1998) also showed that while 20\% of compact groups had an apparent deficiency of CO emission, the rest had CO and FIR properties similar to isolated galaxies. Both Leon et al. (1998) and Verdes-Montenegro et al. (1998) concluded that tidal interactions in compact groups were very important in the groups' evolution.

The aim of this paper is to determine the extent to which the \HI in compact groups has been affected by the environment. This has been investigated previously by Williams \& Rood (1987, hereafter WR87), Huchtmeier (1997, hereafter H97) and Verdes-Montenegro et al. (2001), all of whom observed subsamples of the Hickson Compact Groups (HCGs, Hickson 1982). Their results indicated that the average amount of \HI in the groups was between 40\% and 50\% of the mass expected by summing the mean \HI mass in comparable field galaxies.

Williams \& Rood (1987) compared the \HI content of the compact groups with that of a reference sample of loose groups (RLG), and with a sample of 204 spiral galaxies in the RC2 (de Vaucouleurs et al. 1976). In the south, WR87 observed HCGs 3, 14, 16, 19, 22, 23, 25, 26, 30, 31, 40, 62, 64, 67, 88, 89 and 97. The comparison loose groups were selected to have the same number of galaxies as the compact groups, and to have the same joint distribution of galaxy luminosity, Hubble morphological type. The expected masses therefore were a function of the luminosity of the galaxies. Twenty-seven of the 34 detected compact groups had less \HI than the comparable loose group.

A similar procedure was used by Huchtmeier (1997), using an explicit relation between the integrated blue luminosities of the groups and the expected \HI masses of the groups. The groups observed in the south were HCGs 4, 14, 19, 21, 22, 24, 25, 26, 30, 40, 42, 48, 62, 67, 87, 91 and 97. In comparison to four nearby groups, a reference sample of 146 galaxies (Huchtmeier \& Richter 1988), and to the Virgo cluster spirals, several compact groups appeared to have a very low \HI mass-to-blue luminosity ratio.

These authors did not consider how the \HI mass of the compact groups related to the distribution of \HI masses in the comparison samples. Instead, only the average values of the sample galaxy masses were used. However, any two galaxies with identical optical measurements (eg. diameter, morphology, luminosity) will not necessarily have the same \HI mass. There may be a dependence of the \HI mass on these observables, but the scatter in this correlation is significant.

Verdes-Montenegro et al. (2001) compiled the data from WR87 and H97, and added VLA imaging of 16 HCGs (2, 16, 18, 23, 26, 31, 33, 40, 44, 49, 54, 79, 88, 92, 95, 96). Using all 72 groups from the combined sample, Verdes-Montenegro et al. (2001) calculated the mean \HI ``deficiency'' to be DEF$_{\rm HI} = 0.40\pm0.07$, where DEF$_{\rm HI} \equiv \log[{\rm M(HI)}_{\rm predicted}] - \log[{\rm M(HI)}_{\rm observed}]$. A comparison sample from Haynes \& Giovanelli (1984) was used. This comparison sample is also used in this paper and is described in Section~\ref{estimators}. Calculated values of DEF$_{\rm HI}$ suggested that compact groups contained only 40\% of the \HI mass of comparable field galaxies. Verdes-Montenegro et al. (2001) also deduced that an evolutionary sequence for HCGs could be followed, beginning with \HI being associated with individual galaxies, through to the gas enveloping the entire group, to a group having no detectable \HIn.

This paper compares the \HI content of the galaxies in compact groups with estimated values of \HI mass for field galaxies. The distribution of \HI mass in the reference field galaxies will be specifically included in the comparison. Thus it is possible to identify groups with masses outside the expected range. The upper limits of the non-detected groups are also taken into account using the statistical methods of Buckley \& James (1979).

This paper first deals with the selection of the group sample (Section~\ref{samplesel}), and then with the \HI observations, reduction and the derived parameters (Section~\ref{obsresults}). The mass estimation methods are described in Section~\ref{estimators}, as well as the estimated \HI content of the groups. Section~\ref{discussion} compares the relative contents of the compact groups and the reference sample. For this paper, H$_{0} = 70$ \kms Mpc$^{-1}$.

\section{Sample Selection}\label{samplesel}
Compact groups were selected from the optical catalogues of Hickson (1982, hereafter H82) and Prandoni et al. (1994, hereafter P94). The sample included all compact groups which had a declination $\delta < +2^{\circ}$, and a velocity $cz < 12700$ km s$^{-1}$. A total of 62 groups satisfied these criteria, and are the primary sample of this paper.

HIPASS covers the entire southern sky up to a declination of $\delta<+2^{\circ}$, over the \HI velocity range $-1280 < cz < 12700$ km s$^{-1}$ (Barnes et al. 2001). Using the 13-beam multibeam instrument on the Parkes radio telescope, the survey was completed in March 2000. The velocity resolution of the survey is 18.0 km s$^{-1}$, and the RMS noise per channel is typically 13 mJy.

The Hickson compact groups (H82) - for which a complete catalogue of properties of the constituent galaxies is available (Hickson 1993) - were selected so that a direct comparison with previous results could be made. The Southern compact groups (SCGs, P94) were selected by an automatic routine, and therefore have the potential to be a more complete and less biased sample (P94). However, there has not been a published follow-up survey of the member galaxies, and thus the SCGs do not have complete redshift information.

H82 selected groups from the red Palomar Observatory Sky Survey (POSS) plates by eye, with each of the groups satisfying specific criteria. Firstly, the groups needed to be compact and isolated. Specifically, there needed to be at least 4 galaxies within 3 magnitudes of the brightest galaxy, all within a circle of angular radius $\theta_{\rm G}$, the group radius. The isolation criteria specified that there should be no galaxies within 3 magnitudes of the brightest galaxy within $3\theta_{\rm G}$ of the group centre. The magnitude concordance was used to select physically associated groups without any redshift information, while the isolation criteria ensured that the cores of clusters, and associations in clusters were not identified as compact groups.

P94 selected groups in the same way as H82, except that the search was on COSMOS scans of plates taken with the United Kingdom Schmidt Telescope (UKST), and the search algorithm was machine implemented.

The MIRIAD task {\bf mbspect} was used to inspect the HIPASS spectrum for each group, which was made by averaging over a region of 5$\times$5 pixels, centered on the optically determined position of the group. Each HIPASS pixel is 4 arcminutes on a side. This 20$'\times$20$'$ averaging area ensures that all \HI flux coming from the groups is accounted for, as the largest group in the sample has an optical diameter of $10.8'$.

The spectra were examined by eye for emission lines, and groups with evidence of emission were selected for reobservation. The HCG spectra were searched at each group's known optical velocity (Hickson 1993). The SCGs however, do not have comprehensive velocity information for all their member galaxies, and thus the entire velocity range of HIPASS was examined at each SCG position on the sky.

In total, 19 HCGs and 21 SCGs showed possible \HI emission. Twelve HCGs and ten SCGs were clear non-detections. The primary sample for this paper is shown in Table~\ref{basicdata}. Column (1) gives the name of the compact group, and Columns (2) \& (3) give the position of the group in J2000 coordinates.

\begin{table}
\begin{center}
{\tiny
\begin{tabular}{ccccccc}
\hline\hline
Object & RA & Dec & Integration Time & RMS Noise & Peak Flux & Mass Limit\\
 & (J2000) & (J2000) & (min) & (mJy) & (mJy) & (10$^{9}$ M$_{\odot}$)\\
(1) & (2) & (3) & (4) & (5) & (6) & (7) \\
\hline\hline
HCG 003   & 00:34:27  & $-$07:35:32 & 14 & 6.6  &       & 3.50\\
HCG 004   & 00:34:15  & $-$21:26:47 & 14 & 5.7  & 44.1  & 3.69\\
HCG 006   & 00:39:10  & $-$08:23:43 &    &      &       & 7.43\\
HCG 007   & 00:39:23  & $+$00:52:40 & 14 & 8.4  & 44.5  & 1.14\\
HCG 014   & 01:59:47  & $-$07:01:42 & 14 & 6.9  &       & 1.64\\
HCG 016   & 02:09:33  & $-$10:09:47 & 14 & 6.7  & 117.3 & 0.86\\
HCG 019   & 02:45:45  & $-$12:24:42 & 28 & 7.3  & 24.6  & 1.02\\
HCG 021   & 02:45:17  & $-$17:37:09 & 14 & 10.4 & 21.1  & 2.94\\
HCG 022   & 03:03:31  & $-$15:40:32 & 14 & 7.3  & 67.6  & 0.36\\
HCG 023   & 03:07:06  & $-$09:35:08 & 14 & 7.6  & 36.1  & 1.34\\
HCG 024   & 03:20:18  & $-$10:51:53 &    &      &       & 4.78\\
HCG 025   & 03:20:43  & $-$01:03:07 & 14 & 7.4  & 34.9  & 2.22\\
HCG 026   & 03:21:54  & $-$13:38:45 & 14 & 6.4  & 30.3  & 5.04\\
HCG 028   & 04:27:19  & $-$10:19:00 &    &      &       & 7.27\\
HCG 030   & 04:36:28  & $-$02:49:57 &    &      &       & 1.18\\
HCG 031   & 05:01:36  & $-$04:15:24 & 14 & 6.6  & 170.6 & 0.50\\
HCG 040   & 09:38:54  & $-$04:51:07 &    &      &       & 2.45\\
HCG 042   & 10:00:21  & $-$19:38:57 &    &      &       & 0.93\\
HCG 043   & 10:11:13  & $-$00:01:54 &    &      &       & 5.52\\
HCG 048   & 10:37:45  & $-$27:04:50 & 28 & 7.4  & 19.2  & 0.41\\
HCG 048/1 & 10:37:45  & $-$27:04:50 &    & 7.4  & 18.9  \\
HCG 048/2 & 10:37:45  & $-$27:04:50 &    & 7.4  & 19.2  \\
HCG 062   & 12:53:08  & $-$09:13:27 &    &      &       & 1.06\\
HCG 063   & 13:02:08  & $-$32:46:04 & 14 & 7.0  & 30.0  & 5.35\\
HCG 064   & 13:25:43  & $-$03:51:28 &    &      &       & 6.55\\
HCG 067   & 13:49:03  & $-$07:12:20 & 14 & 7.4  & 23.4  & 3.19\\
HCG 086   & 18:47:05  & $-$30:49:33 & 14 & 5.6  &       & 2.08\\
HCG 087   & 20:48:11  & $-$19:50:35 &    &      &       & 4.43\\
HCG 088   & 20:52:22  & $-$05:45:28 & 14 & 12.0 & 52.2  & 2.14\\
HCG 089   & 21:20:10  & $-$03:54:31 & 14 & 14.0 & 71.8  & 4.59\\
HCG 090   & 22:02:05  & $-$31:58:00 &    &      &       & 0.40\\
HCG 091   & 22:09:10  & $-$27:47:45 & 14 & 20.2 &       & 3.05\\
HCG 097   & 23:47:22  & $-$02:19:34 &    &      &       & 2.59\\
HCG 098   & 23:54:12  & $+$00:22:24 &    &      &       & 3.68\\
\hline
SCG 06    & 01:49:10  & $-$27:06:42 & 14 & 6.3  &       \\
SCG 07    & 00:37:43  & $-$33:41:24 & 28 & 4.8  & 33.0  & 4.62\\
SCG 14    & 01:56:28  & $-$20:05:51 & 14 & 6.8  &       \\
SCG 15    & 00:32:41  & $-$25:36:47 & 14 & 7.1  & 40.4  & 0.50\\
SCG 17    & 00:51:14  & $-$32:25:59 & 14 & 8.7  &       \\
SCG 18    & 01:18:32  & $-$17:02:24 & 14 & 6.4  &       \\
SCG 20    & 00:14:56  & $-$24:05:25 & 70 & 7.5  & 39.0  & 3.38\\
SCG 24    & 00:19:14  & $-$26:42:49 & 28 & 5.4  &       \\
SCG 25    & 02:53:11  & $-$09:29:12 & 14 & 6.8  &       \\
SCG 27    & 03:03:07  & $-$22:12:10 & 14 & 7.3  &       \\
SCG 28    & 00:47:32  & $-$21:29:16 & 14 & 5.8  & 37.6  & 2.41\\
SCG 33    & 03:04:51  & $-$12:04:17 & 28 & 6.2  &       \\
SCG 34    & 01:19:27  & $-$17:25:16 & 14 & 6.5  &       \\
SCG 35    & 03:37:06  & $-$07:41:39 & 14 & 6.1  &       \\
SCG 39    & 00:37:09  & $-$34:57.49 & 28 & 5.1  &       \\
SCG 43    & 01:15:57  & $-$29:46:00 & 14 & 6.7  & 34.7  & 5.29\\
SCG 49    & 02:55:08  & $-$21:35:43 & 14 & 6.1  &       \\
SCG 51    & 00:33:36  & $-$27:47:00 & 14 & 6.5  & 132.2 & 0.14\\
SCG 54    & 00:05:58  & $-$36:06:54 & 14 & 6.4  &       \\
SCG 55    & 00:58:58  & $-$28:17:38 & 14 & 5.3  &       \\
SCG 57    & 01:49:57  & $-$27:48:32 & 14 & 10.7 &       \\
\hline\hline
\end{tabular}
\caption{Basic observational data for the compact group sample and pointed \HI observations using the MX mode. Descriptions of the quantities in each column are given in the text.}\label{basicdata}
}
\end{center}
\end{table}

\section{HI Data}\label{obsresults}
\subsection{Follow-up Observations}
To confirm the \HI properties of these groups, and to more accurately measure their masses, the 40 compact groups detected in HIPASS were reobserved with the Parkes Telescope Multibeam instrument during November 2001, April 2002, and January 2003. These observations were carried out using the narrowband mode of the correlator (MX mode: 8 MHz bandwidth, 1024 channels), giving a velocity resolution of 2.0 km s$^{-1}$. Each group was observed for a total on-source integration time of 14 minutes, using the MX observing mode.

The MX observing mode places the central 7 beams of the Multibeam on-source, in turn, for 2 minutes each. The 12 minutes that each beam spends off-source is used to measure the bandpass for subtraction from the on-source observation. The final spectrum is made by combining the 7 separate bandpass-subtracted spectra.

The FWHM of each of the beams is $\sim14'$, meaning each of the compact groups is entirely covered by the beam, when the beam is placed at the group centre. The largest group is HCG 21 with a diameter of $10.8'$, while the smallest groups are SCGs 54 \& 57, with a diameter of $0.8'$. The median group diameter is $2.5'$. The diameter of the groups is defined by the smallest circle which contains all the galaxies. The telescope was pointed at the centres of these circles.

\subsection{Data Reduction}
Initial reduction of the data used {\bf LiveData} (Barnes 1998), which subtracts the bandpass estimates from the on-source observations, and {\bf gridzilla} which then combines the seven MX observations into one. The reduced spectra are shown in Figure~\ref{detspectra} for the detected compact groups. The velocity shown in the spectra in Figure~\ref{detspectra} is optical velocity, in the heliocentric frame.

The compact group spectra were analysed and parameterised using the MIRIAD task {\bf mmspect}, an extended version of {\bf mbspect}. This task examines the combined spectra from the MX observations and calculates an estimate of the continuum baseline level using a Gaussian smooth, which can then be subtracted.  The baseline fit is shown in Figure~\ref{detspectra} as a solid line. The width of the fitting gaussian was always $>40$ channels to ensure that small scale fluctuations do not affect the baseline calculation. The sources of the baseline instability after bandpass subtraction in {\bf LiveData} are solar interference which changes rapidly, and pointings near a strong continuum source which creates standing waves between the dish surface and the receiver.

The channels within the profile window (marked in Figure~\ref{detspectra} as dashed lines) were excluded from the baseline fit, and the baseline has been interpolated across this window. The profile window was determined so that the sides of the window cut the baseline at the edges of the profile, and within the window the profile does not cross the baseline.

It is assumed that the \HI in each group is concentrated at the centre of the observation, and that it is optically thin. This makes the \HI masses presented here lower limits. The \HI mass of a galaxy is given by:
\begin{equation}\label{himass}
{\rm M}_{\rm HI} = 2.36\times10^{5} \left (\frac{v^{2}}{\rm H_{0}^{2}}\right ) {\rm S}_{\rm int} \mbox{ M}_{\odot}
\end{equation}
where $v$ is the recessional velocity of the galaxy, and ${\rm S}_{\rm int}$ is the integrated flux from the galaxy.
If a correction for the location of the \HI in the beam were to be applied to these observations, the maximum correction needed can be calculated if all the \HI is considered to be at the edge of the group diameter, and thus farthest away from the centre of the beam as possible. Since the largest group diameter is $10.8'$, and the FWHM of the observing beams is 14.3$'$, the maximum possible flux correction needed would be 48\% upwards.

Seven groups have had multiple observations combined to produce the final spectrum. These observations were combined using the miriad task {\bf imcomb}. Two observations were combined for HCG 19, HCG 48, SCG 7, SCG 24, SCG 33 and SCG 39, and five observations were combined for SCG 20.

The RMS noise level of the spectrum has been calculated from the non-profile channels. This calculation is performed before the baseline fit has been subtracted, so it includes the effect of the baseline ripple. It is much harder to find a signal superimposed on a varying baseline than it is on a flat one, and this measure quantifies this difference. The RMS noise level is given in Table~\ref{basicdata}.

\begin{table}
\begin{center}
{\tiny
\begin{tabular}{ccccccc}
\hline\hline
Object & Integrated Flux & Mean Velocity & Mean Velocity & 50\% Width & 20\% Width & Opt Group\\
 &           & Heliocentric  & GSR    &        &        & Vel Disp \\
 & (Jy \kms) & (\kms) & (\kms) & (\kms) & (\kms) & (\kms)\\
(1) & (2) & (3) & (4) & (5) & (6) & (7)\\
\hline\hline
HCG 003   &                 &      & \dw\nw7883\pw\nw\nw &                &                 & 251.2    \\
HCG 004   & \nw3.2$\pm$0.1  & 8058 & \nw8092$\pm$\nw2    & \nw83$\pm$\nw3 & 111$\pm$\nw\nw5 & 338.9    \\
HCG 006   &                 &      & \dw11480\pw\nw\nw   &                &                 & 251.2    \\
HCG 007   & \nw7.6$\pm$0.6  & 4404 & \nw4506$\pm$12      & 222$\pm$24     & 297$\pm$\nw38   & \nw89.1  \\
HCG 014   &                 &      & \dw\nw5393\pw\nw\nw &                &              & 331.1    \\
HCG 016   & 26.5$\pm$0.5    & 3888 & \nw3900$\pm$\nw5    & 228$\pm$11     & 369$\pm$\nw17   & 123.0    \\
HCG 019   & \nw3.4$\pm$0.3  & 4286 & \nw4252$\pm$\nw9    & 167$\pm$19     & 213$\pm$\nw29   &          \\
HCG 021   & \nw2.5$\pm$0.6  & 7262 & \nw7228$\pm$\nw6    & 189$\pm$11     & 201$\pm$\nw18   & 112.2    \\
HCG 022   & \nw8.0$\pm$0.2  & 2574 & \nw2533$\pm$\nw1    & 137$\pm$\nw3   & 160$\pm$\nw\nw5 & \nw43.7  \\
HCG 023   & \nw7.7$\pm$1.0  & 4901 & \nw4875$\pm$25      & 335$\pm$50     & 429$\pm$\nw79   & 169.8    \\
HCG 024   &                 &      & \dw\nw9209\pw\nw\nw &                &                 & 199.5    \\
HCG 025   & \nw7.0$\pm$0.7  & 6285 & \nw6273$\pm$\nw9    & 283$\pm$18     & 323$\pm$\nw28   & \nw61.7  \\
HCG 026   & \nw9.0$\pm$1.2  & 9506 & \nw9458$\pm$\nw9    & 440$\pm$18     & 471$\pm$\nw28   & 199.5    \\
HCG 028   &                 &      & \dw11357\pw\nw\nw   &                &                 & \nw85.1  \\
HCG 030   &                 &      & \nw4577\pw\nw\nw    &                &                 & \nw72.4  \\
HCG 031   & 23.9$\pm$0.3    & 4057 & \nw3968$\pm$\nw2    & 125$\pm$\nw5   & 222$\pm$\nw\nw7 & \nw85.1  \\
HCG 040   &                 &      & \dw\nw6599\pw\nw\nw &                &                 & 147.9    \\
HCG 042   &                 &      & \dw\nw4069\pw\nw\nw &                &                 & 213.8    \\
HCG 043   &                 &      & \dw\nw9893\pw\nw\nw &                &                 & 223.9    \\
HCG 048   & \nw5.1$\pm$1.7  &      & \nw2707$\pm$24      & 454$\pm$77     & 514$\pm$116     & 302.0    \\
HCG 048/1 & \nw3.4$\pm$0.8  & 2497 & \nw2571$\pm$22      & 290$\pm$45     & 337$\pm$\nw70   &          \\
HCG 048/2 & \nw1.7$\pm$0.4  & 2787 & \nw2861$\pm$\nw5    & 164$\pm$11     & 177$\pm$\nw16   &          \\
HCG 062   &                 &      & \dw\nw4327\pw\nw\nw &                &                 & 288.4    \\
HCG 063   & \nw5.0$\pm$0.6  & 9312 & \dw\nw9479$\pm$11   & 260$\pm$21     & 301$\pm$\nw33   & 131.8    \\
HCG 064   &                 &      & \dw10785\pw\nw\nw   &                &                 & 213.8    \\
HCG 067   & \nw3.5$\pm$0.4  & 7731 & \nw7865$\pm$23      & 186$\pm$47     & 285$\pm$\nw73   & 208.9    \\
HCG 086   &                 &      & \dw\nw6073\pw\nw\nw &                &                 & 269.2    \\
HCG 087   &                 &      & \dw\nw8868\pw\nw\nw &                &                 & 120.2    \\
HCG 088   & \nw5.9$\pm$0.7  & 6015 & \nw6161$\pm$\nw4    & 168$\pm$\nw8   & 188$\pm$\nw12   & \nw26.9  \\
HCG 089   & \nw8.7$\pm$0.6  & 8872 & \nw9023$\pm$\nw8    & 154$\pm$16     & 211$\pm$\nw24   & \nw55.0  \\
HCG 090   &                 &      & \dw\nw2672\pw\nw\nw &                &                 & 100.0    \\
HCG 091   &                 &      & \dw\nw7353\pw\nw\nw &                &                 & 182.0    \\
HCG 097   &                 &      & \dw\nw6785\pw\nw\nw &                &                 & 371.5    \\
HCG 098   &                 &      & \dw\nw8083\pw\nw\nw &                &                 & 120.2    \\
\hline
SCG 06 & & & & \\
SCG 07 & 13.4$\pm$1.5 & 8988 & \nw9054$\pm$26 & 619$\pm$51 & 724$\pm$80 \\
SCG 14 & & & & \\
SCG 15 & \nw7.5$\pm$0.7 & 2934 & \nw2982$\pm$\nw4 & 270$\pm$\nw9 & 293$\pm$13 \\
SCG 17 & & & & \\
SCG 18 & & & & \\
SCG 20 & \nw9.4$\pm$1.1 & 7680 & \nw7740$\pm$19 & 358$\pm$38 & 438$\pm$60 \\
SCG 24 & & & & \\
SCG 25 & & & & \\
SCG 27 & & & & \\
SCG 28 & \nw3.9$\pm$0.2 & 6479 & \nw6540$\pm$\nw8 & 139$\pm$16 & 211$\pm$24 \\
SCG 33 & & & & \\
SCG 34 & & & & \\
SCG 35 & & & & \\
SCG 39 & & & & \\
SCG 43 & \nw1.8$\pm$0.1 & 9654 & \nw9691$\pm$\nw3 & 60$\pm$\nw6 & 100$\pm$10 \\
SCG 49 & & & & \\
SCG 51 & 21.6$\pm$0.2 & 1539 & \nw1581$\pm$\nw4 & 110$\pm$\nw7 & 341$\pm$11 \\
SCG 54 & & & & \\
SCG 55 & & & & \\
SCG 57 & & & & \\
\hline\hline
\end{tabular}
\caption{Measured \HI parameters for the compact groups detected by the pointed \HI observations. Descriptions of the quantities in each column are given in the text.}\label{hidata}
}
\end{center}
\end{table}

\begin{figure}
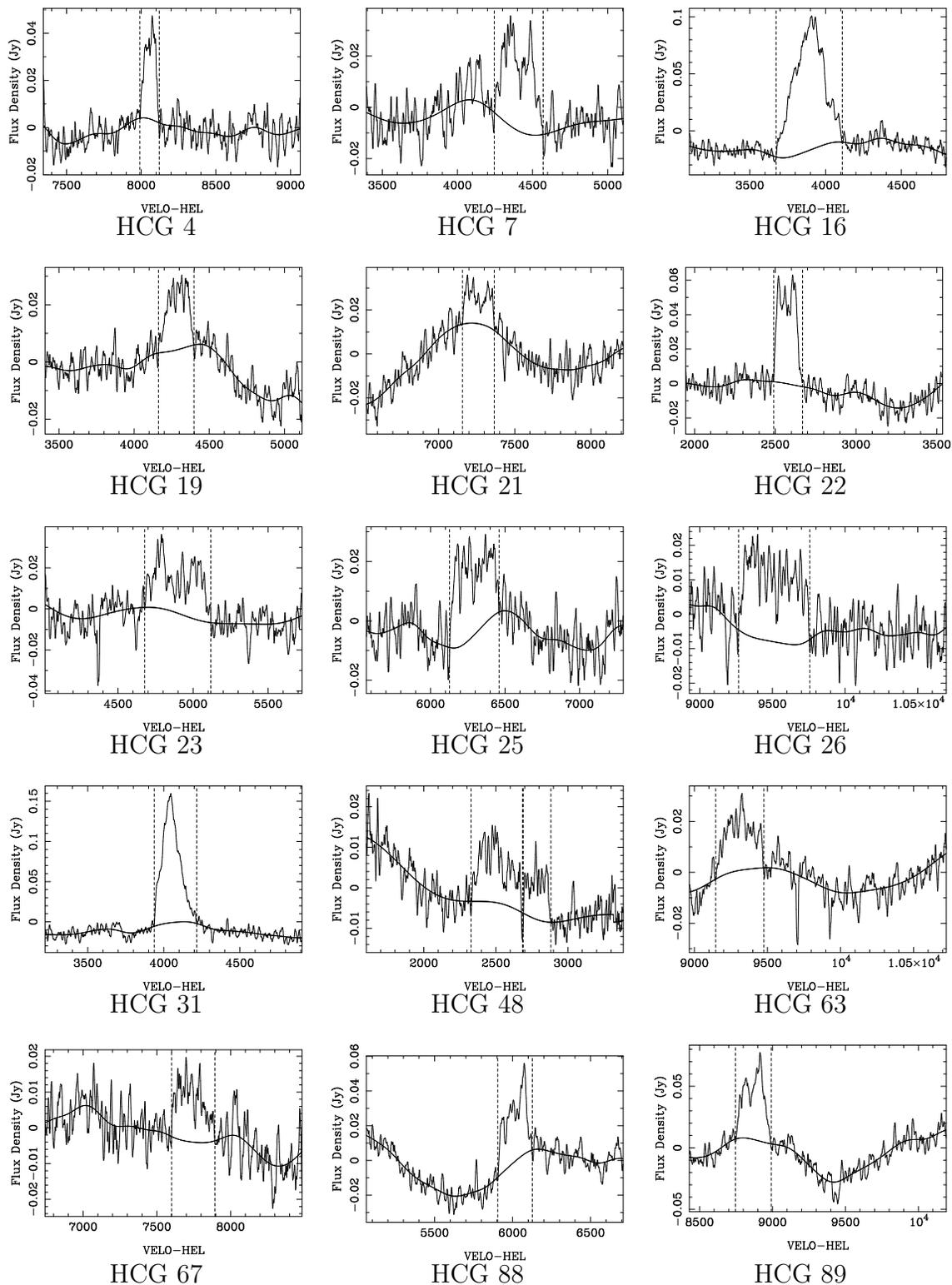

\begin{tabular}{ccc}
\includegraphics[scale=0.21,angle=270]{groupspectra/detected/HCG004.ps} & \includegraphics[scale=0.21,angle=270]{groupspectra/detected/HCG007.ps} & \includegraphics[scale=0.21,angle=270]{groupspectra/detected/HCG016.ps} \\
HCG 4 & HCG 7 & HCG 16\\
\includegraphics[scale=0.21,angle=270]{groupspectra/detected/HCG019.ps} & \includegraphics[scale=0.21,angle=270]{groupspectra/detected/HCG021.ps} & \includegraphics[scale=0.21,angle=270]{groupspectra/detected/HCG022.ps} \\
HCG 19 & HCG 21 & HCG 22\\
\includegraphics[scale=0.21,angle=270]{groupspectra/detected/HCG023.ps} & \includegraphics[scale=0.21,angle=270]{groupspectra/detected/HCG025.ps} & \includegraphics[scale=0.21,angle=270]{groupspectra/detected/HCG026.ps} \\
HCG 23 & HCG 25 & HCG 26\\
\includegraphics[scale=0.21,angle=270]{groupspectra/detected/HCG031.ps} & \includegraphics[scale=0.21,angle=270]{groupspectra/detected/HCG048.ps} & \includegraphics[scale=0.21,angle=270]{groupspectra/detected/HCG063.ps} \\
HCG 31 & HCG 48 & HCG 63\\
\includegraphics[scale=0.21,angle=270]{groupspectra/detected/HCG067.ps} & \includegraphics[scale=0.21,angle=270]{groupspectra/detected/HCG088.ps} & \includegraphics[scale=0.21,angle=270]{groupspectra/detected/HCG089.ps} \\
HCG 67 & HCG 88 & HCG 89 \\
\end{tabular}
\caption{Narrowband MX spectra of the detected compact groups. The solid line in each spectrum is the estimated baseline, and the vertical dashed lines surround the emission region. The emission from HCG 48 has been split into two domains, separated by the area where the profile crosses the baseline, to provide a reliable estimate of the flux.}\label{detspectra}
\end{figure}
\addtocounter{figure}{-1}
\begin{figure}
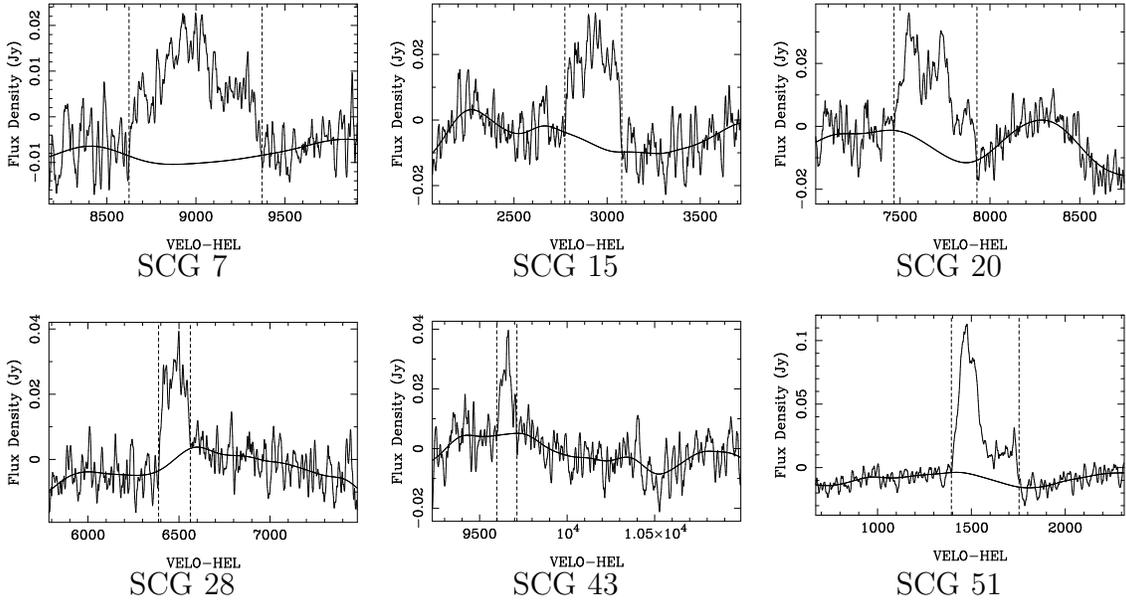

\begin{tabular}{ccc}
\includegraphics[scale=0.21,angle=270]{groupspectra/detected/SCG07.ps} & \includegraphics[scale=0.21,angle=270]{groupspectra/detected/SCG15.ps} & \includegraphics[scale=0.21,angle=270]{groupspectra/detected/SCG20.ps} \\
SCG 7 & SCG 15 & SCG 20\\
\includegraphics[scale=0.21,angle=270]{groupspectra/detected/SCG28.ps} & \includegraphics[scale=0.21,angle=270]{groupspectra/detected/SCG43.ps} & \includegraphics[scale=0.21,angle=270]{groupspectra/detected/SCG51.ps} \\
SCG 28 & SCG 43 & SCG 51 
\end{tabular}
\caption{({\it cont.}) Narrowband spectra of the detected compact groups.}
\end{figure}

\subsection{HI Observational Parameters}

Table~\ref{basicdata} gives the basic \HI observational data for the compact group sample. Column (4) gives the total narrowband integration time on the group in minutes. Column (5) gives the RMS noise level (per smoothed channel) of the narrowband observation (mJy). If data in this column is not listed, the group has not been detected in the HIPASS data, and was not reobserved. Column (6) gives the observed peak flux (mJy) of the \HI emission from the group. If data in this column is not listed, the group has not been detected by the narrowband observations. Column (7) gives the observable \HI mass limit per HIPASS spectral channel, obtained using the limit given in Section~\ref{hipasslimits}, and the known velocity of the group (Hickson 1993). For the SCGs, this value is calculated using the observed velocity. If an SCG has not been detected, this column is left blank.

Table~\ref{hidata} gives the \HI profile data for each of the groups. Column (1) lists the group name. Column (2) gives the integrated flux of the detected profile in Jy km s$^{-1}$. Column (3) gives the observed intensity-weighted mean velocity of the profile (heliocentric), computed in the following way:
\begin{equation}\label{meanvel}
\overline{v} = \frac{\sum I\times v}{\sum I}
\end{equation}
where $\overline{v}$ is the mean velocity of the profile, and $I$ is the intensity of the emission at velocity $v$. Column (4) shows this velocity after correction to the Galactic Standard of Rest (GSR) frame, as used by Braun \& Burton (1999):
\begin{eqnarray}
\overline{v}_{\rm GSR} & = & \overline{v} + 9\cos(l)\cos(b)+12\sin(l)\cos(b)-7\sin(b)\nonumber\\
& & +220\sin(l)\cos(b)\label{lgsr}
\end{eqnarray}
where $l$ \& $b$ are the galactic latitude and longitude of the group respectively, and $\overline{v}$ is in the heliocentric frame. The uncertainties of columns (3) and (4) are the same, but only column (4) is shown with errors because the GSR velocities will be used from now on. Column (5) gives the maximum 50\% width of the emission profile, and column (6) is the maximum 20\% width. Maximum widths are determined from the two points at the appropriate fraction of the peak flux farthest away from each other in the emission profile. Column (7) gives the optical radial velocity dispersion of the group, which is the RMS of the galaxy velocities with respect to the velocity centroid (Hickson 1993).

\begin{figure}
\begin{center}
\begin{tabular}{cc}
\includegraphics[scale=0.35]{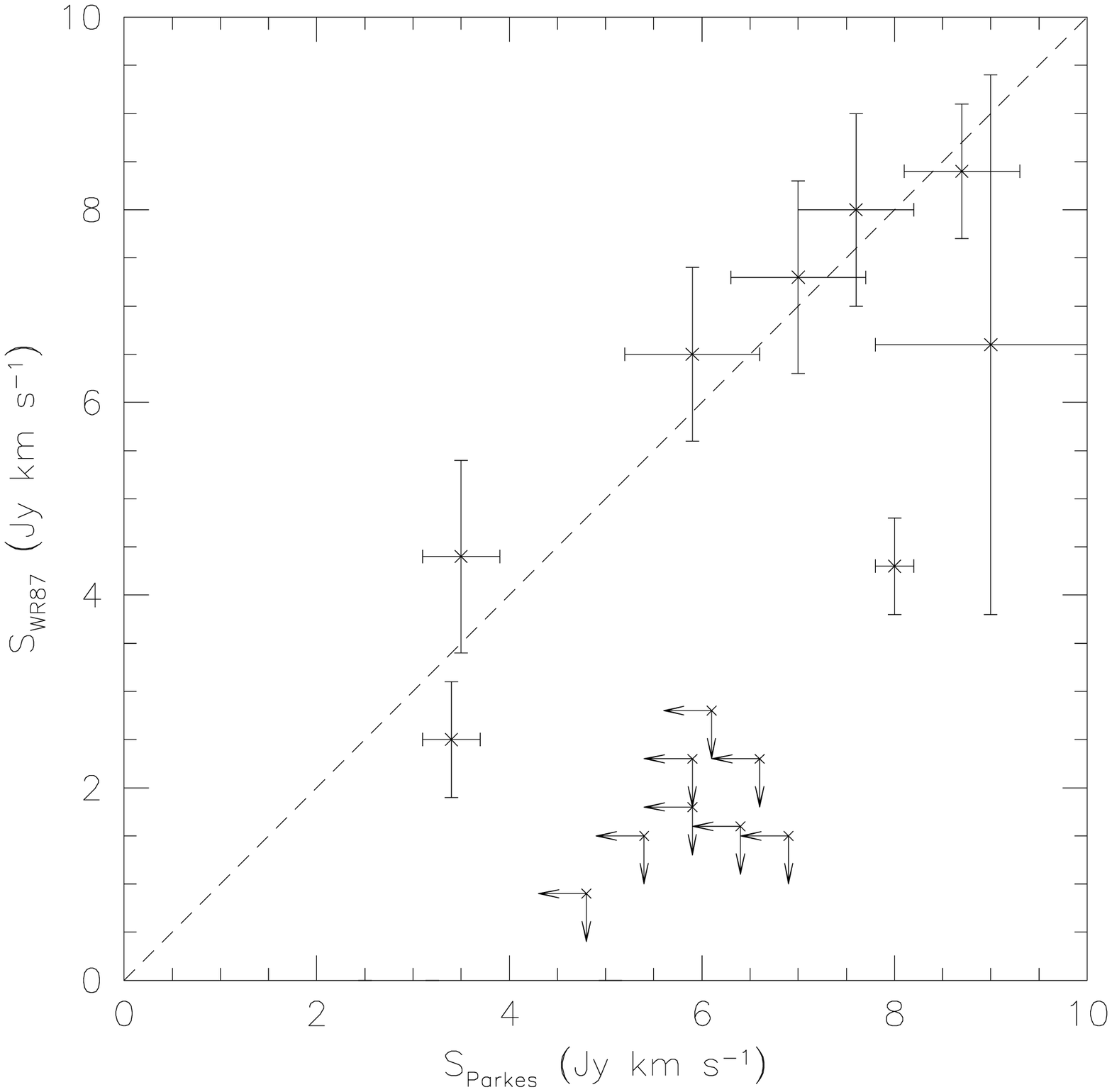} & \includegraphics[scale=0.35]{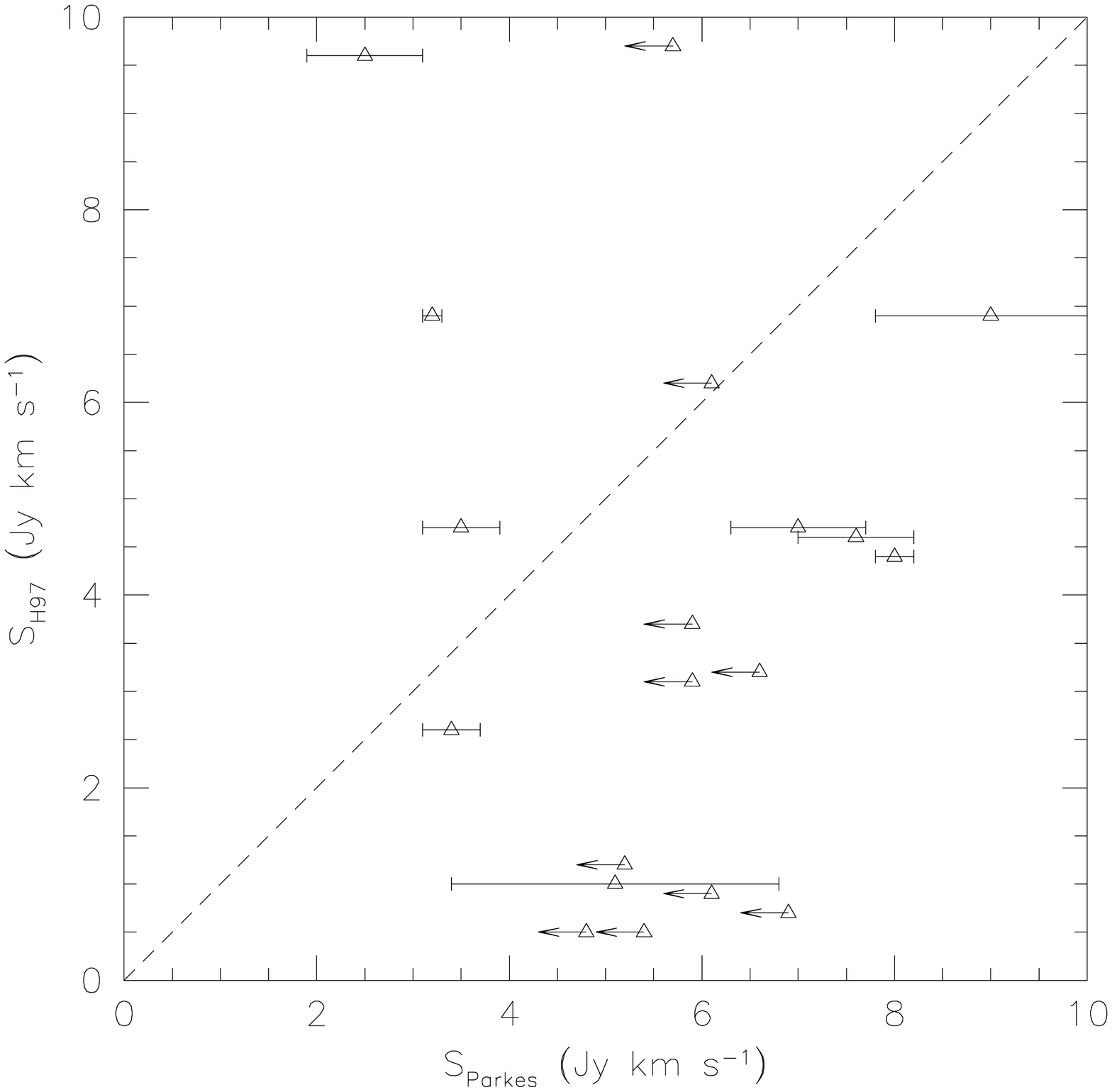}\\
(a)&(b)
\end{tabular}
\includegraphics[scale=0.35]{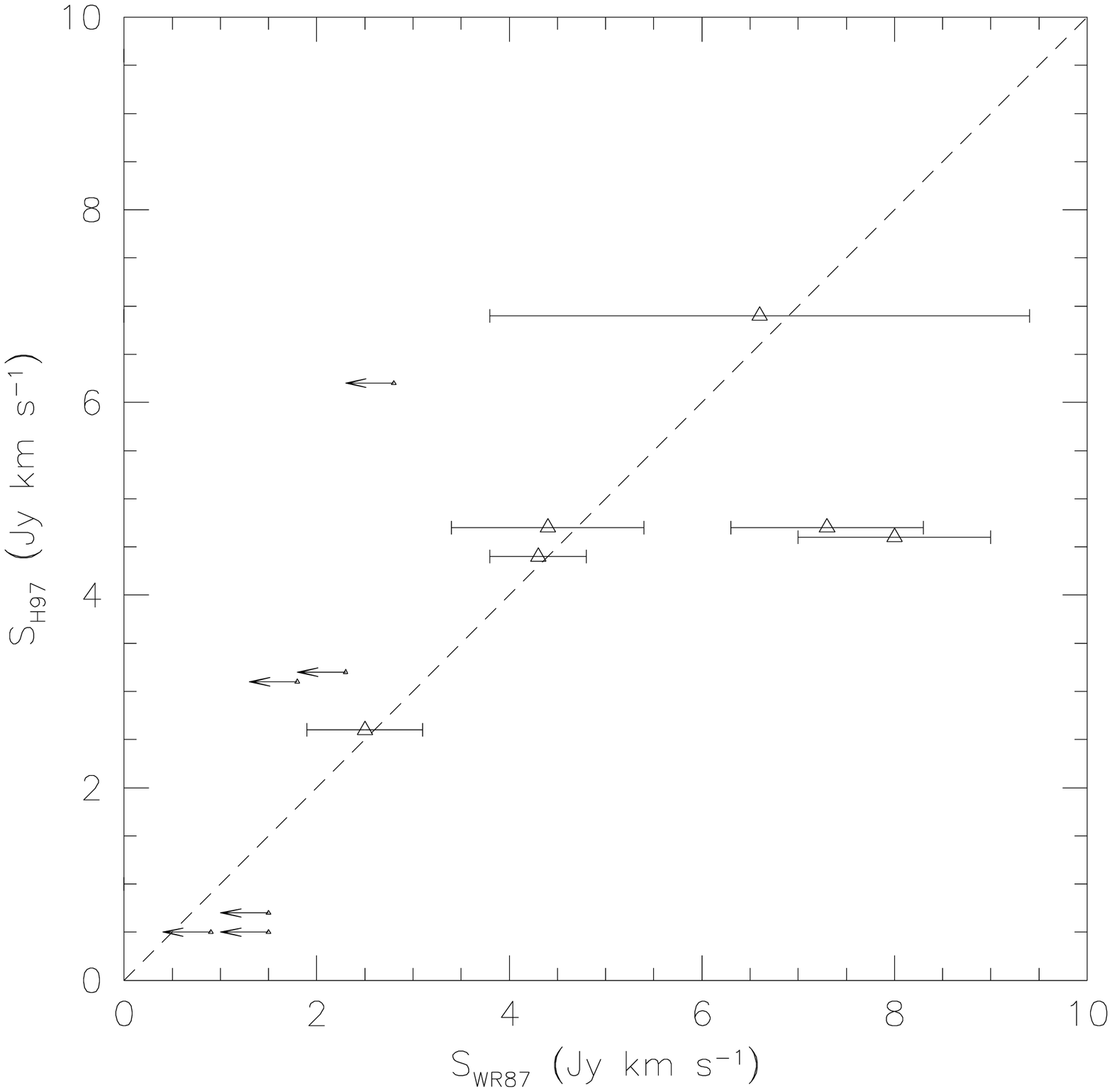}\\
(c)
\caption{The fluxes of the compact groups studied in this sample are compared to the measurements of (a) WR87 and (b) H97, and the fluxes measured by WR87 and H97 are compared to each other in (c). The line of equality is shown as a dashed line. No flux errors or upper limits were listed in H97. Upper flux limits are indicated by arrows. Three WR87 detections are not shown on these plots; HCG 16, HCG 23, and HCG 31, which lie above the axis limits.}\label{comparison}
\end{center}
\end{figure}

A comparison of the \HI fluxes measured by this study, WR87 and H97 is shown in Figure~\ref{comparison}. Only the Hickson compact groups listed in this paper are compared in Figure~\ref{comparison}. Good agreement is seen with the fluxes of WR87, while for H97, there is no correlation. Between H97 and WR87, there are four groups which agree well - HCGs 19, 22, 26 and 67. Three of these groups (HCG 19, 26, 67) have a reasonable agreement to the narrowband observations, within the WR87 uncertainties. WR87 used the NRAO 91m and the Arecibo 305m telescopes for their observations, while H97 used the Effelsberg 100m telescope. There is no clear reason why the agreement to the observations of WR87 is better than to those from H97. The upper limits found by HIPASS are generally higher than those determined by WR87 and H97, as expected.

\subsubsection{Errors in measured parameters}
The uncertainties for the 50\% profile width and the 20\% width are given by the semi-empirical relation found by Schneider et al. (1986):
\begin{equation}\label{uncwidthorig}
\Delta v(f) = 1.5\sqrt{2}\mbox{ }|v_{80}-v_{20}|\left(\frac{0.25}{f(1-f)}\right)({\rm S/N})^{-1}
\end{equation}
where $\Delta v(f)$ is the uncertainty in the profile width at $(100f)\%$ of the peak flux, (ie. the 50\% width has $f=0.50$), $v_{80}$ is the 80\% width of the profile, $v_{20}$ is the 20\% width, and (S/N) is the signal-to-noise ratio of the spectrum. 

This can be estimated by the following equation:
\begin{equation}\label{signaltonoise}
({\rm S/N}) \equiv \frac{{\rm S}_{\rm int}}{\Delta{\rm S}_{\rm int}} = \frac{{\rm S}_{\rm int}}{\sigma v_{50}}\sqrt{\frac{v_{50}}{13.2}}
\end{equation}
 where ${\rm S}_{\rm int}$ is the integrated flux of the profile, $\sigma$ is the RMS noise level, $v_{50}$ is the 50\% width of the profile, and the reduced velocity resolution of the narrowband observations is 13.2 \kms. The inverse of equation~\ref{signaltonoise} can be used as the uncertainty estimate of the integrated flux.

It is necessary to change Equation~\ref{uncwidthorig} to take the 50\% width, instead of the 80\% width. Thus $v_{80}$ is replaced by $v_{50}$, and the factor of 3.0 is twice the original value because $|v_{80}-v_{20}|$ is two times $|v_{50}-v_{20}|$, assuming the profile sides are straight. Equation~\ref{uncwidthorig} therefore becomes:
\begin{equation}\label{uncwidth}
\Delta v(f) = 3.0\sqrt{2}\mbox{ }|v_{50}-v_{20}|\left(\frac{0.25}{f(1-f)}\right)({\rm S/N})^{-1}
\end{equation}
The error on the mean velocity can be estimated in a similar way (Schneider et al. 1986), and is given by:
\begin{equation}\label{uncmean}
\Delta\overline{v} = \frac{3.0}{\sqrt{2}}\mbox{ }|v_{50}-v_{20}|({\rm S/N})^{-1}
\end{equation}
where $\Delta\overline{v}$ is the uncertainty on the intensity-weighted mean velocity.

\subsubsection{HIPASS Detection Limits}\label{hipasslimits}
The detection of an object depends not only on its flux, but also on the number of channels over which the flux is spread. The detection experiment of the group sample was done using HIPASS data, thus to find the upper limits for the non-detected groups, knowledge of the HIPASS detection limit is needed. This can be deduced from the integrated flux ${\rm S}_{\rm int}$ and the 50\% velocity width $v_{50}$ of the 4082 galaxies in the HIPASS catalogue (Meyer et al. 2004). For a detection, the signal-to-noise ratio over the width of the galaxy profile needs to be greater than a specified level, which is taken as the detection limit. The mean single-channel RMS noise level of HIPASS is $\sigma=13$ mJy.

In the HIPASS catalogue, most galaxies are detected with SNR $>5\sigma$, and there is a sharp dropoff in detections at lower signal-to-noise. The approximate HIPASS detection limit is thus:
\begin{equation}
{\rm S}_{\rm int} \gtrsim 5\frac{\sigma}{\sqrt{\rm n}}v_{50}
\end{equation}
where ${\rm n}$ is the number of uncorrelated channels defining the profile. An estimate of ${\rm n}$ is $v_{50}/18.0$ since the velocity resolution of HIPASS is 18.0 km s$^{-1}$. Here any channel-to-channel correlations are ignored, and normal statistics is assumed.

The upper mass limit for the non-detected galaxies is then:
\begin{equation}\label{upperlimit}
{\rm M}_{\rm HI} \lesssim (2.36\times10^{5}{\rm D}^{2})(5\sigma\sqrt{18.0})\sqrt{v_{50}} \mbox{ M}_{\odot}
\end{equation}
where ${\rm D}$ is the distance to the group in Mpc, calculated using the median of the corrected optical velocities of the group members.

To estimate the upper \HI mass limits for the non-detected groups requires that an estimate of the total velocity width of the group be made. If the inclinations of each galaxy and their optical rotation speeds were known for all the groups, and the galaxies which were likely to have \HI could be identified, then estimating $v_{50}$ would be straight-forward. However, the only related quantity available for the HCGs is the optical velocity dispersion.

\begin{figure}[h]
\begin{center}
\includegraphics[scale=0.4]{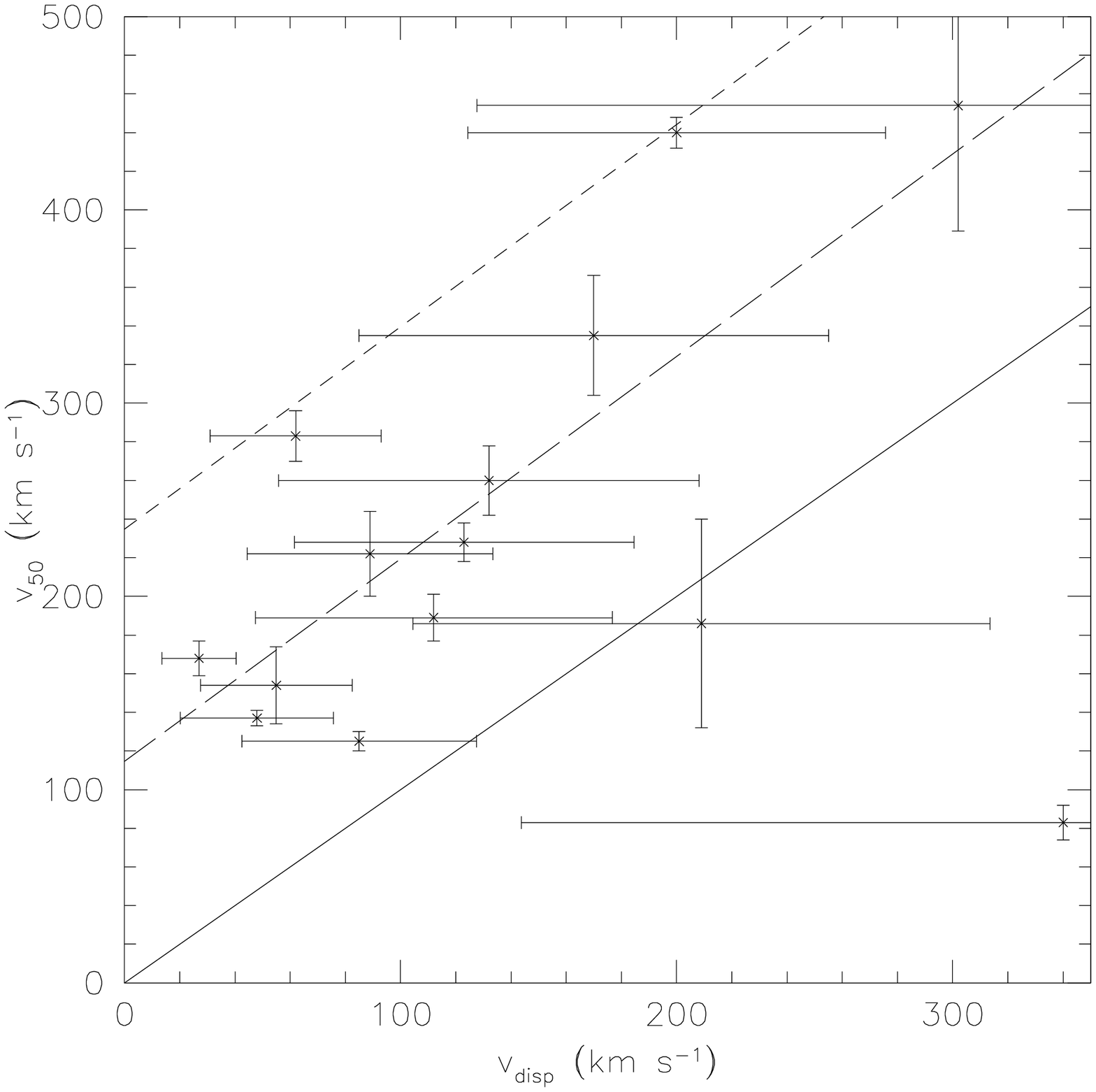}
\caption{The 50\% \HI velocity width $v_{50}$ is plotted against the optical velocity dispersion of the group for the detected compact groups. The long-dashed line is the least squares fit to the data, the short-dashed line is the fit offset to become an upper limit, and the solid line is $v_{50} = v_{\rm disp}$.}\label{dispwidth}
\end{center}
\end{figure}

A plot of $v_{50}$ versus the optical radial velocity dispersion $v_{\rm disp}$ for each of the detected compact groups is shown in Figure~\ref{dispwidth}. It shows that $v_{\rm disp}$ is in practice, a useful lower limit on the \HI velocity width. This is expected if the group is unresolved spectrally in \HIn. To get the upper mass limit requires a upper limit on $v_{50}$. This limit is obtained by calculating the least squares fit to the data in Figure~\ref{dispwidth} and shifting it upwards to make it an upper bound. The group HCG 4 (lower right corner) has been left out of the regression calculation as it clearly does not fit with the general trend. It is conjectured that in HCG 4, only the central face-on spiral HCG004a is detected in \HIn, as the other two members of the group are ellipticals.

The equation for the line which represents a reasonable upper limit on the width of the \HI profile, given the optical velocity dispersion is:
\begin{equation}\label{maxwidth}
v_{50}  =  1.05v_{\rm disp} + 230 \mbox{ km s}^{-1}
\end{equation}

The upper mass limits for the non-detected HCGs have been calculated using equation~\ref{maxwidth}, and are shown in Column (2) of Table~\ref{content}.

\section{Predicted HI Masses}\label{estimators}
The null hypothesis is that the \HI detected in a compact group is the sum of the \HI in the individual member galaxies, if they were field galaxies with the same optical properties.

A test of the null hypothesis requires a reliable method of calculating a galaxy's likely \HI mass, depending on its type and size. In this section, three mass estimation methods are described.

Two of the methods - the Mass-Diameter relation, and the Mass-Luminosity relation - are described by Haynes \& Giovanelli (1984, hereafter HG84), who examine a sample of isolated field galaxies to investigate whether correlations exist between each galaxy's \HI mass, and its optical properties. The correlations with the optical diameter of a galaxy, and with the blue luminosity are also examined.

The isolated galaxy sample of HG84 was extracted from the Catalogue of Isolated Galaxies (CIG) by Karachentseva (1973). This catalogue contained 1052 galaxies with a magnitude limit of +15.7 in the POSS red prints. The galaxies in the catalogue were determined to be isolated on the following basis. If a galaxy of diameter $d$ has a neighbour galaxy of diameter $d_{1}$, with the constraint $\frac{1}{4}d\leq d_{1}\leq4d$, then the first galaxy is isolated from the second if it lies more than 20$d_{1}$ away from it. The CIG contains galaxies which are isolated from all other galaxies according to this metric.

The HG84 sample contains only those galaxies from the CIG which are also included in the Uppsala General Catalog of Galaxies (UGC, Nilson 1973), and lie at a declination which can be observed by the Arecibo 305m telescope ($-1^{\circ}<\delta<+38^{\circ}$). This leaves a sample of 324 isolated galaxies.

The third mass estimation method is that used by WR87, and H97, and is a simplified version of the mass-luminosity relation of HG84.

The application of each of these methods to this sample is discussed in turn.

\subsection{Optical Diameter as an indicator of HI mass}

HG84 report a strong correlation between the optical major diameter of a galaxy and the galaxy's \HI mass, which takes the form:
\begin{equation}\label{massdiamfull}
\log(h^{2}\expect{{\rm M}_{\rm HI}}) = c_{1}(t) + c_{2}(t)\log(h{\rm D}_{\rm l})^2
\end{equation}
where $\expect{{\rm M}_{\rm HI}}$ is the expected \HI mass (M$_{\odot}$), $c_{1}$ \& $c_{2}$ are scalar parameters which depend on the galaxy's numerical morphological type $t$, D$_{\rm l}$ is the length of the major axis of the galaxy (kpc), and $h$ is the Hubble factor such that ${\rm H}_{0} = 100h$ km s$^{-1}$ Mpc$^{-1}$. The numerical morphological type ranges from 0 to 10, with later type galaxies having a higher numerical type (Sandage \& Tammann 1981). That is, ellipticals have a type $t=0$, and peculiar galaxies are $t=10$. Table~\ref{haynesmdcoeff} lists the numerical types associated with galaxy morphology. This system differs from that of the definition given in the RC2 (de Vaucouleurs, 1976), which has $t=-5$ for ellipticals, $t=-2$ for S0, $t=0$ for S0a, and increases to $t=10$ for irregular galaxies. The system of Sandage \& Tammann (1981) is used here because it matches with the system used by HG84.

Equation~\ref{massdiamfull} reduces to:
\begin{equation}\label{massdiam}
\expect{{\rm M}_{\rm HI}} = \frac{1}{h^{2}}10^{c_{1}(t)}(h{\rm D}_{\rm l})^{2c_{2}(t)}
\end{equation}

Equation~\ref{massdiam} states that a galaxy has a well-defined surface density of \HI dependent on its morphological type, which is expressed in the constant $10^{c_{1}(t)}$. However, there is an additional dependence on the diameter of the galaxy, and this is expressed by the power $c_{2}(t)$.

HG84 determined the values of $c_{1}(t)$ and $c_{2}(t)$ for the various morphological types from the isolated sample of 324 galaxies after observations with the Arecibo 305m Telescope, and the Green Bank 91m Telescope. The rate of non-detections for the morphological types 0, 1 \& 2 was very high in HG84, with only 14 of 30 galaxies of these types detected. The coefficients were calculated based only on the detected galaxies, so they are biased towards higher mass galaxies. To remove this bias, the results of Chamaraux et al. (1986, hereafter C86) are used here for galaxies of type 1 and 2. For elliptical galaxies (numerical type 0), the \HI mass is assumed to be zero, as only 6\% of RC3 ellipticals have been detected in HIPASS (Sadler et al. 2002). The low \HI detection rate of ellipticals makes it impossible to reliably estimate the range of \HI masses that these galaxies would have.

The $c_{1}$ \& $c_{2}$ values are shown in Table~\ref{haynesmdcoeff}, along with the standard error of the estimate. The standard error of the estimate (s.e.e.) is a measure of how far away a data point is from its predicted value, and is defined as:
\begin{equation}\label{see}
{\rm s.e.e.} \equiv \sqrt{\frac{\Sigma (y - y')^{2}}{{\rm N}}}
\end{equation}
where $y$ is the known value of some data point, $y'$ is the value predicted by some function which fits the data point, and N is the number of data points that the function is fitted to.

\begin{table}
\begin{center}
\begin{tabular}{cccccc}
\hline\hline
Type $t$ & Morphology & $c_{1}$ & $c_{2}$ & s.e.e. & Source\\
\hline\hline
1 & S0 & 5.68 & 1.00 & 0.25 & C86\\
2 & S0a & 6.21 & 1.00 & 0.10 & C86\\
3,4 & Sa,Sab & 7.17 & 0.82 & 0.31 & HG84\\
5 & Sb & 7.29 & 0.83 & 0.25 & HG84\\
6 & Sbc & 7.27 & 0.85 & 0.17 & HG84\\
7 & Sc & 6.91 & 0.95 & 0.18 & HG84\\
8,9 & Scd,Sd,Irr,Sm, & 7.00 & 0.94 & 0.17 & HG84\\
 & Sdm,dwarf sp \\
10 & Pec & 7.75 & 0.66 & 0.19 & HG84\\
\hline\hline
\end{tabular}
\caption{Coefficients of the Mass-Diameter relation (MDR) from Haynes \& Giovanelli (1984) and Chamaraux et al. (1986).}\label{haynesmdcoeff}
\end{center}
\end{table}

C86 take into account the upper limits of detection, using the method of Chamaraux (1987). This method uses the mean \HI surface density, similar to HG84, and thus the method for determining $\expect{{\rm M}_{\rm HI}}$ is the same.
The values found by C86 are lower than those found by HG84 for the same morphological types. There was no attempt by C86 to determine the residual dependence on the galaxy's diameter, so $c_{2}=1$ for these types.

\subsection{Blue Luminosity as an indicator of HI mass}\label{lumsection}

The distance-independent relationship between the \HI mass of a galaxy and its blue luminosity ${\rm L}_{\rm B}$ can be expressed as:
\begin{equation}\label{lumdiamfull}
\log(h^{2}\expect{{\rm M}_{\rm HI}}) = c_{3}(t) + c_{4}(t)\log(h^{2}{\rm L}_{\rm B})
\end{equation}
which, like the relation between diameter and mass, can be written as:
\begin{equation}\label{lumdiam}
\expect{{\rm M}_{\rm HI}} = \frac{1}{h^{2}}10^{c_{3}(t)}(h^{2}{\rm L}_{\rm B})^{c_{4}(t)}
\end{equation}

\begin{table}
\begin{center}
\begin{tabular}{ccccc}
\hline\hline
Type $t$ & $c_{3}$ & $c_{4}$ & s.e.e. & Source\\
\hline\hline
1,2 & 2.93 & 0.63 & 0.36 & HG84\\
3,4 & 2.87 & 0.66 & 0.38 & HG84\\
5 & 2.99 & 0.66 & 0.36 & HG84\\
6 & 2.77 & 0.69 & 0.29 & HG84\\
7 & 3.03 & 0.66 & 0.26 & HG84\\
8,9 & 1.40 & 0.84 & 0.30 & HG84\\
10 & 4.99 & 0.45 & 0.25 & HG84\\
\hline\hline
\end{tabular}
\caption{Coefficients of the Mass-Luminosity relation (MLR) from Haynes \& Giovanelli (1984).}\label{haynesmlcoeff}
\end{center}
\end{table}

Table~\ref{haynesmlcoeff} gives the values for the constants in Equation~\ref{lumdiam}. We can see that, in comparison to the MDR, the errors involved with the MLR are slightly higher, ie. on average there is a greater dispersion in the relationship between the predicted and measured values of \HI mass when using Equation~\ref{lumdiam}. The MLR was not investigated by C86, thus all values in Table~\ref{haynesmlcoeff} are from HG84. As HG84 again did not take into account the non-detection rates for galaxies with early type morphology, the values quoted for types $t=1,2$ are most likely overestimated.

\subsection{The Simplified Mass-Luminosity Relation}
A simplified version of the MLR was used by both WR87 and H97. For galaxies with type $t=0,1,2$, H97 quotes the values $\expect{{\rm M}_{\rm HI}}/{\rm L}_{\rm B} = 0.03$, and $\expect{{\rm M}_{\rm HI}}/{\rm L}_{\rm B} = 0.65$ for types $t>2$. Since dependence on luminosity was not considered, $c_{4}=1$, and thus equation~\ref{lumdiam} becomes $\expect{{\rm M}_{\rm HI}}/{\rm L}_{\rm B} = 10^{c_{3}(t)}$. Thus $c_{3} = -1.52$ for galaxies with types $t=1$ \& $2$ (early types), and $c_{3} = -0.19$ for later types. This relation will be referred to as the Simplified Mass-Luminosity relation (SMLR).

\subsection{Compact Group HI Content}\label{groupcontent}

For each of the HCGs, Hickson (1993) has classified the members by galaxy type. The expected \HI content of each group was calculated by summing the expected M$_{\rm HI}$ for each member's morphological type. This calculation is repeated here for each of the three methods described above.

Table~\ref{content} shows the \HI masses calculated with each of the methods described above, as well as the observed \HI mass. Column (2) of Table~\ref{content} gives the \HI mass of the compact group. If the group has not been detected in these observations, an upper mass limit is given. This limit is obtained using the HIPASS detection limit, which is detailed in Section~\ref{hipasslimits}. It would be inappropriate to use the limits from the follow-up survey, because the {\em detection experiment} was done using the HIPASS data. Columns (3), (4) and (5) of Table~\ref{content} give the expected masses of the groups calculated using the MDR, MLR and SMLR respectively. The errors on these estimates are calculated from the quoted standard errors listed for each of the methods.

\begin{table*}
\begin{center}
{\tiny
\begin{tabular}{ccccc}
\hline\hline
Object & HI Mass & Expected Mass (MDR) & Expected Mass (MLR) & Expected Mass (SMLR)\\
 & (10$^{9}$ M$_{\odot}$) & (10$^{9}$ M$_{\odot}$) & (10$^{9}$ M$_{\odot}$) & (10$^{9}$ M$_{\odot}$)\\
(1) & (2) & (3) & (4) & (5)\\
\hline\hline
HCG 003 & \nw\nw\dw\nw$\leq$18.3 & \nw5.7$\pm$\nw0.8 & 10.7$\pm$\nw2.0 & \nw8.4$\pm$\nw3.9\\
HCG 004 & 10.1$\pm$\nw0.3 & 14.5$\pm$\nw2.5 & 23.3$\pm$\nw6.8 & 48.3$\pm$24.1\\
HCG 006 & \nw\nw\dw\nw$\leq$38.9 & \nw7.9$\pm$\nw1.7 & \nw7.9$\pm$\nw1.7 & 22.6$\pm$\nw9.1\\
HCG 007 & \nw7.4$\pm$\nw0.6 & 13.3$\pm$\nw2.1 & 22.3$\pm$\nw4.0 & 34.2$\pm$11.8\\
HCG 014 & \nw\nw\dw\nw$\leq$\nw9.3 & \nw8.3$\pm$\nw1.4 & \nw7.3$\pm$\nw2.1 & \nw9.8$\pm$\nw4.4\\
HCG 016 & 19.4$\pm$\nw0.4 & \nw6.3$\pm$\nw1.3 & \nw8.7$\pm$\nw2.1 & 36.8$\pm$13.8\\
HCG 019 & \nw3.0$\pm$\nw0.3 & \nw5.9$\pm$\nw0.8 & \nw3.8$\pm$\nw0.7 & \nw5.1$\pm$\nw1.7\\
HCG 021 & \nw6.3$\pm$\nw1.4 & 27.1$\pm$\nw5.3 & 14.3$\pm$\nw3.2 & 67.0$\pm$22.9\\
HCG 022 & \nw2.5$\pm$\nw0.1 & \nw0.1$\pm$\nw0.0 & \nw0.4$\pm$\nw0.1 & \nw0.1$\pm$\nw0.0\\
HCG 023 & \nw8.8$\pm$\nw1.1 & \nw6.8$\pm$\nw1.5 & \nw4.1$\pm$\nw0.8 & \nw6.9$\pm$\nw2.7\\
HCG 024 & \nw\nw\dw\nw$\leq$23.6 & \nw1.9$\pm$\nw0.5 & \nw5.1$\pm$\nw1.5 & 19.9$\pm$\nw9.8\\
HCG 025 & 13.3$\pm$\nw1.4 & \nw9.6$\pm$\nw1.5 & \nw7.3$\pm$\nw1.4 & 20.8$\pm$\nw7.3\\
HCG 026 & 38.8$\pm$\nw5.3 & \nw9.6$\pm$\nw1.3 & \nw7.1$\pm$\nw1.1 & 11.1$\pm$\nw3.5\\
HCG 028 & \nw\nw\dw\nw$\leq$30.6 & 10.6$\pm$\nw2.6 & \nw7.2$\pm$\nw2.2 & \nw9.7$\pm$\nw4.7\\
HCG 030 & \nw\nw\dw\nw$\leq$\nw4.9 & \nw2.1$\pm$\nw0.3 & \nw4.6$\pm$\nw0.9 & \nw3.3$\pm$\nw1.2\\
HCG 031 & 18.1$\pm$\nw0.2 & \nw2.1$\pm$\nw0.3 & \nw7.9$\pm$\nw2.0 & 15.9$\pm$\nw6.8\\
HCG 040 & \nw\nw\dw\nw$\leq$11.3 & \nw7.9$\pm$\nw1.5 & \nw7.0$\pm$\nw2.0 & \nw8.2$\pm$\nw3.8\\
HCG 042 & \nw\nw\dw\nw$\leq$4.7 & \nw0.4$\pm$\nw0.1 & \nw1.7$\pm$\nw0.5 & \nw0.5$\pm$\nw0.2\\
HCG 043 & \nw\nw\dw\nw$\leq$28.0 & 19.4$\pm$\nw2.5 & 22.3$\pm$\nw4.5 & 42.2$\pm$13.5\\
HCG 048 & \nw1.8$\pm$\nw0.4 & \nw0.6$\pm$\nw0.1 & \nw1.0$\pm$\nw0.2 & \nw1.1$\pm$\nw0.6\\
HCG 048/1 & \nw1.1$\pm$\nw0.2 &  &  & \\
HCG 048/2 & \nw0.7$\pm$\nw0.1 &  &  & \\
HCG 062 & \nw\nw\dw\nw$\leq$\nw5.7 & \nw1.7$\pm$\nw0.4 & \nw3.4$\pm$\nw0.9 & \nw1.3$\pm$\nw0.5\\
HCG 063 & 21.6$\pm$\nw2.6 & 17.7$\pm$\nw2.4 & 14.2$\pm$\nw2.8 & 48.1$\pm$18.1\\
HCG 064 & \nw\nw\dw\nw$\leq$32.9 & 24.7$\pm$\nw3.1 & 14.0$\pm$\nw2.6 & 44.0$\pm$17.1\\
HCG 067 & 10.4$\pm$\nw1.2 & 26.2$\pm$\nw5.7 & 11.0$\pm$\nw2.6 & 17.2$\pm$\nw6.2\\
HCG 086 & \nw\nw\dw\nw$\leq$11.1 & \nw0.4$\pm$\nw0.1 & \nw1.4$\pm$\nw0.4 & \nw0.3$\pm$\nw0.1\\
HCG 087 & \nw\nw\dw\nw$\leq$19.7 & \nw8.8$\pm$\nw1.5 & \nw7.9$\pm$\nw1.9 & \nw7.5$\pm$\nw3.2\\
HCG 088 & 10.8$\pm$\nw1.0 & 25.0$\pm$\nw3.0 & 18.2$\pm$\nw3.8 & 47.1$\pm$13.2\\
HCG 089 & 34.1$\pm$2.2 & 13.4$\pm$\nw1.4 & 10.7$\pm$\nw1.5 & 25.1$\pm$\nw7.1\\
HCG 090 & \nw\nw\dw\nw$\leq$\nw1.7 & \nw7.7$\pm$\nw1.7 & \nw6.4$\pm$\nw1.7 & 30.4$\pm$10.9\\
HCG 091 & \nw\nw\dw\nw$\leq$14.7 & 33.2$\pm$\nw4.1 & 17.9$\pm$\nw3.4 & 60.3$\pm$23.3\\
HCG 097 & \nw\nw\dw\nw$\leq$15.2 & \nw8.2$\pm$\nw1.1 & \nw7.0$\pm$\nw1.4 & 12.7$\pm$\nw5.8\\
HCG 098 & \nw\nw\dw\nw$\leq$16.4 & \nw2.0$\pm$\nw0.5 & \nw5.0$\pm$\nw1.4 & \nw2.5$\pm$\nw1.1\\
\hline
SCG 06 &  &  &  & \\
SCG 07 & 52.9$\pm$\nw6.1 & 19.7$\pm$\nw7.1 & \nw5.4$\pm$\nw8.2 & \nw4.8$\pm$\nw1.8\\
SCG 14 &  &  &  & \\
SCG 15 & \nw3.2$\pm$\nw0.3 & \nw0.7$\pm$\nw0.2 & \nw2.1$\pm$\nw0.2 & \nw0.3$\pm$\nw0.1\\
SCG 17 &  &  &  & \\
SCG 18 &  &  &  & \\
SCG 20 & 27.1$\pm$\nw3.1 & \nw5.6$\pm$\nw2.9 & \nw2.1$\pm$\nw3.3 & 12.5$\pm$\nw6.2\\
SCG 24 &  &  &  & \\
SCG 25 &  &  &  & \\
SCG 27 &  &  &  & \\
SCG 28 & \nw8.0$\pm$\nw0.4 & \nw6.7$\pm$\nw1.9 & \nw4.4$\pm$\nw2.5 & \nw3.7$\pm$\nw1.1\\
SCG 33 &  &  &  & \\
SCG 34 &  &  &  & \\
SCG 35 &  &  &  & \\
SCG 39 &  &  &  & \\
SCG 43 & \nw8.1$\pm$\nw0.3 & \nw7.8$\pm$\nw3.4 & \nw5.3$\pm$\nw3.8 & 15.5$\pm$\nw7.1\\
SCG 49 &  &  &  & \\
SCG 51 & \nw2.6$\pm$\nw0.1 &  &  & \\
SCG 54 &  &  &  & \\
SCG 55 &  &  &  & \\
\hline\hline
\end{tabular}
\caption{Predicted \HI masses for each compact group using the phenomenological relationships of HG84 and H97. Details of each of the mass estimates are described in the text.}\label{content}
}
\end{center}
\end{table*}

\subsubsection{HI content of the compact groups}
\begin{figure*}
\begin{center}
\begin{tabular}{cc}
\includegraphics[scale=0.38]{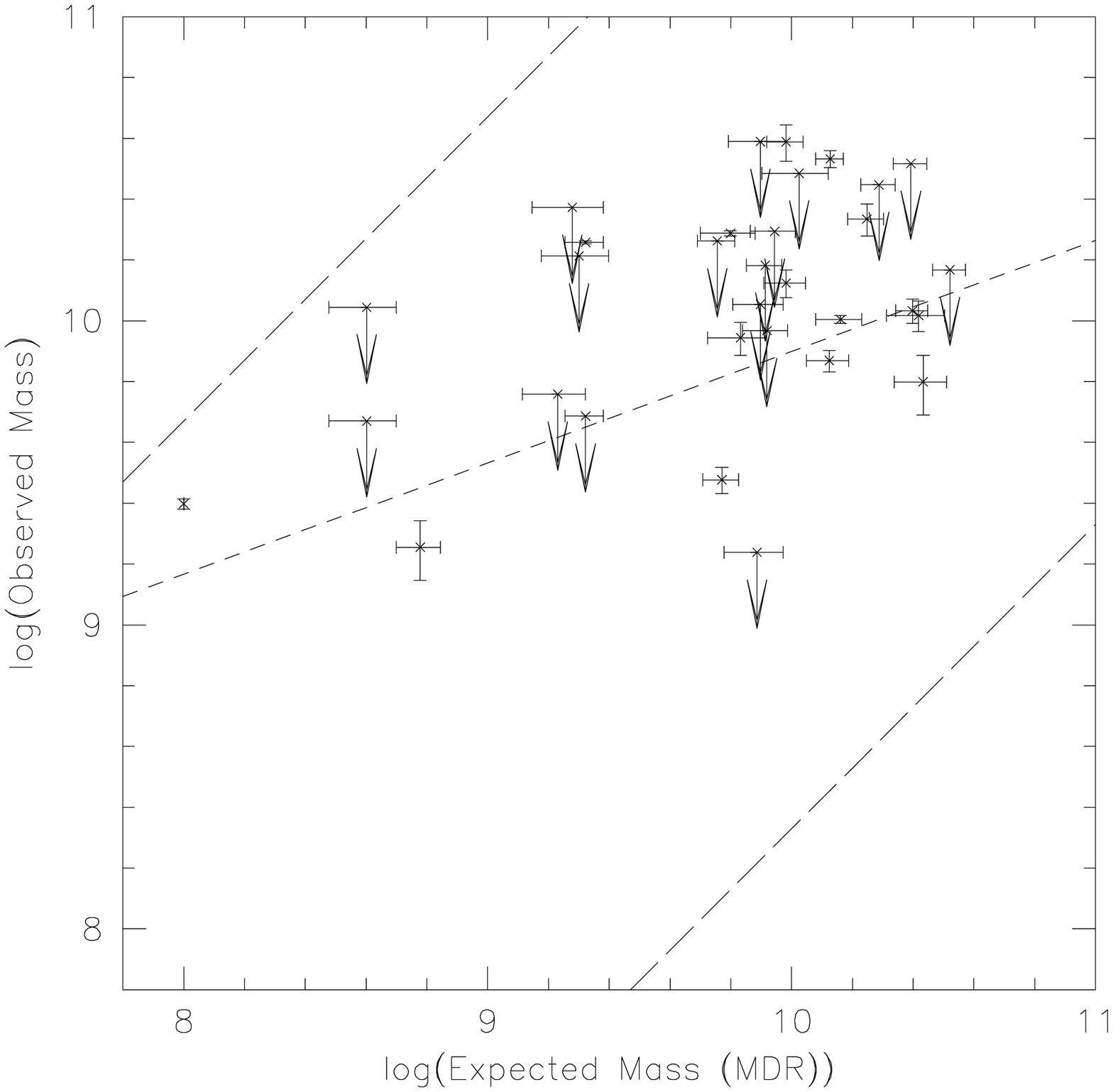} & \includegraphics[scale=0.38]{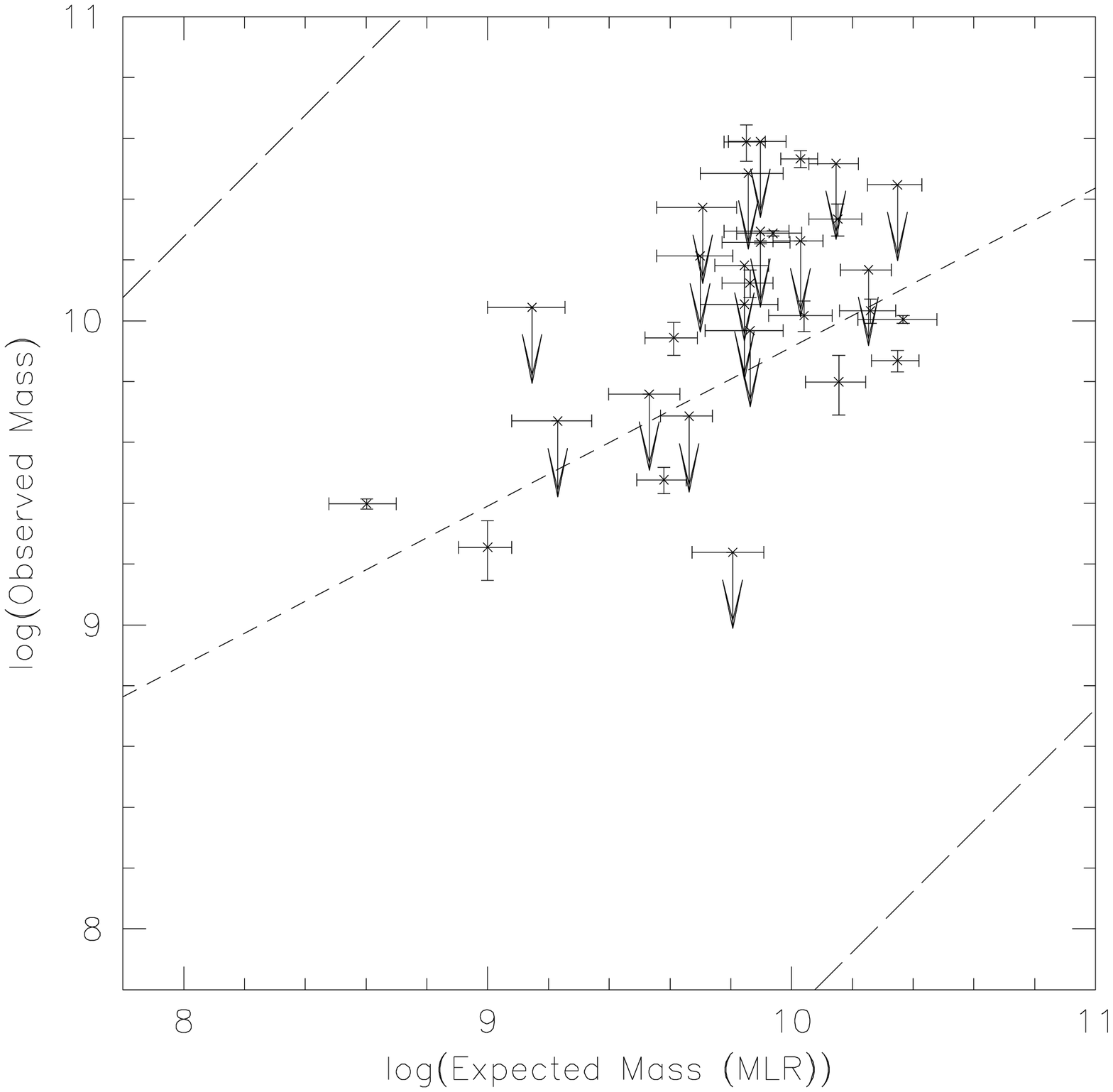} \\
(a) & (b) \\
\end{tabular}
\includegraphics[scale=0.38]{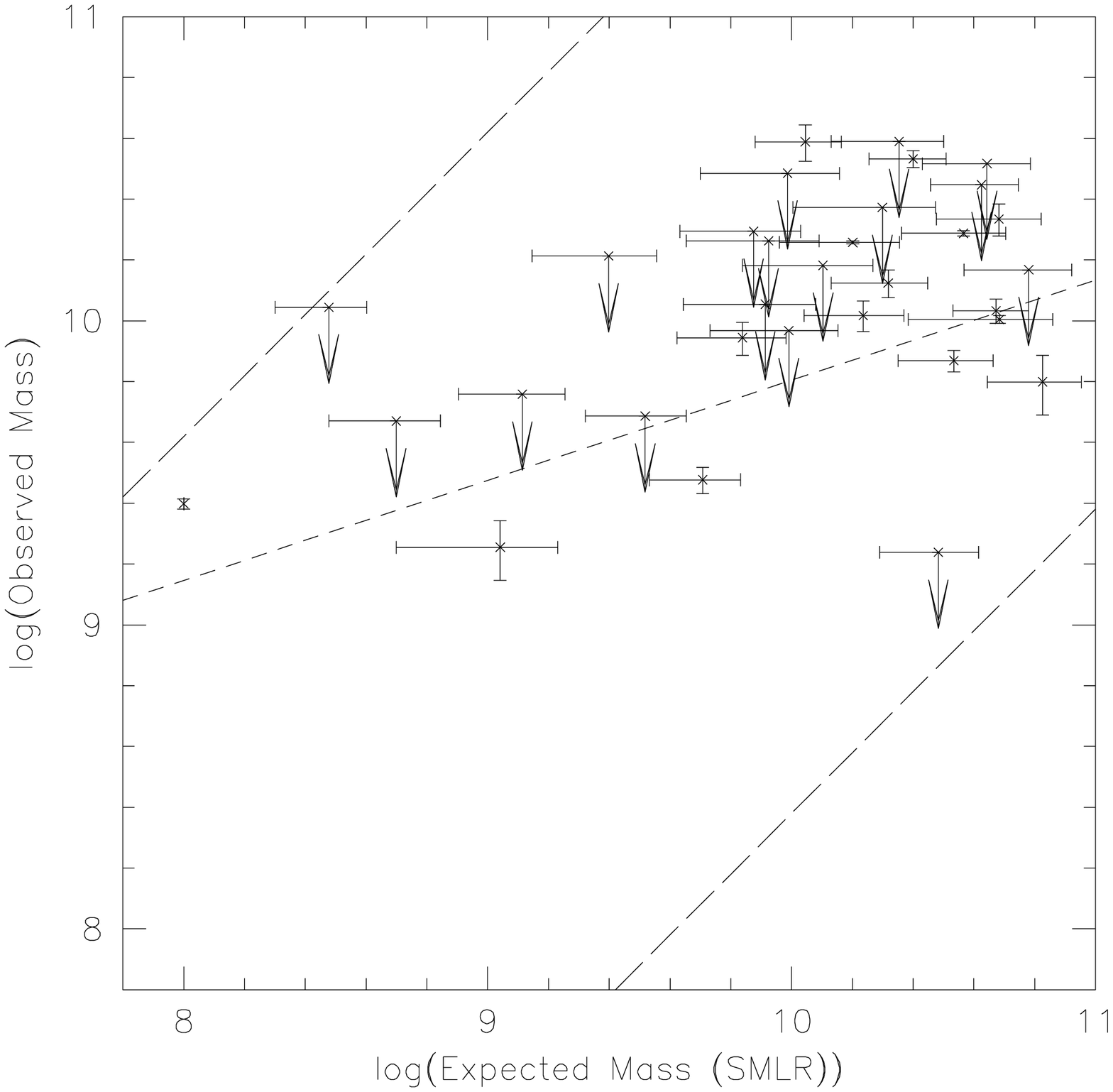} \\
(c)
\caption{The observed mass is plotted against the expected mass for the (a) MDR, (b) MLR, (c) SMLR. The long-dashed line shows the $3\sigma$ errors on each of the estimation techniques, as found by HG84 and H97, and are described further in the text. The short-dashed line is the Buckley-James fit to the data, including the non-detected groups. Upper limits for non-detections are indicated.}\label{expectplots}
\end{center}
\end{figure*}

The plot of expected mass vs observed mass for the HCGs is given in Figure~\ref{expectplots} for each of the estimation methods described above. We no longer include the SCG detections, because without full knowledge of the optical properties of the sample as a whole, including them would only add noise to the analysis. Each plot in Figure~\ref{expectplots} has an area bounded by long-dashed lines. These lines represent the 3$\sigma$ errors for each of the estimation techniques. For a galaxy whose expected \HI mass was estimated using the MDR technique for example, there is a range of observable masses for that galaxy which could be considered consistent with that estimate, because the spread of the observed galaxy masses in HG84 is broad. The long-dashed lines in Figure~\ref{expectplots} show the maximum and minimum observable \HI masses consistent with each expected mass, averaged over all morphological types. For Figure~\ref{expectplots} (a) \& (b), the mass uncertainties are obtained from HG84, and for (c) from H97.

The traditional way of assessing the relative \HI content of a sample of galaxies to the field is to use the deficiency parameter, which is defined in the following way:
\begin{equation}\label{defdef}
{\rm DEF}_{\rm HI}\equiv\log[{\rm M(HI)}_{\rm pred}]-\log[{\rm M(HI)}_{\rm obs}]
\end{equation}
where M(HI)$_{\rm pred}$ is the predicted average \HI mass of a field galaxy with the same optical properties as the sample galaxy, and M(HI)$_{\rm obs}$ is the observed \HI mass of the galaxy. If DEF$_{\rm HI}\leq0$ then the galaxy would not be considered deficient, while a DEF$_{\rm HI}>0$ would indicate a \HI deficiency. The M(HI)$_{\rm pred}$ has an uncertainty associated with it due to the natural spread of masses seen in the field, so the deficiency parameter will also have an uncertainty.

Verdes-Montenegro et al. (2001) calculated the average deficiency of their sample of 50 HCGs to be $\expect{{\rm DEF}_{\rm HI}}=0.40\pm0.07$. 
They considered a group to have an anomalous \HI content when the amount predicted varied from the amount observed by twice the mean error on the predicted mass, otherwise the \HI content is considered to be normal. In this way, only 23 of these 50 groups could be considered deficient, and these 23 groups have an average deficieny of 0.73. Of the other 27 groups, 3 have an anomalously high \HI mass, and the other 24 are in the range considered normal by Verdes-Montenegro et al. (2001). Thus an average measure of deficiency does not give an indication of the properties of the sample as a whole. A further 14 groups were not detected by Verdes-Montenegro et al. (2001) and could not be included in the analysis.

To properly measure the correlation between the expected mass and the observed mass, a technique that takes into account the upper limits of the data is required. The Buckley-James method (Buckley \& James 1979; Isobe, Fiegelson \& Nelson 1986) uses the Kaplan-Meier estimator and the EM algorithm (Expectation, Maximization) to find a linear regression fit to a data set which has upper limits. The EM method determines a fit using estimates of the values of the upper limits (censored points), and changes these estimates iteratively to maximise the likelihood estimators for the unknown parameters. To estimate the values of the censored points, knowledge of the distribution around the regression lines is needed. The Kaplan-Meier estimator uses the known values for the uncensored data to estimate this distribution, and hence give values to the censored points. Simulations by Buckley \& James (1979) showed that this estimator performed well even if 50\% of the points were censored unevenly along the distribution, finding the true value of the slope with a small number ($n=20$) of points.

The fits for each of the methods are shown in Table~\ref{buckleyjames}, and the coefficients are for the relation in equation~\ref{fiteq}:
\begin{equation}\label{fiteq}
\log\expect{{\rm M}_{\rm HI}} = {\rm a}\log{\rm M}_{\rm HI}+{\rm b}
\end{equation}
The estimate of the scatter $\sigma$ is given by the relation:
\begin{equation}\label{sigma}
\sigma = \sqrt{\sum_{i=1}^{n}\frac{(y_{i}-\expect{y_{i}})^{2}}{n-2}}
\end{equation}
where $y_{i}$ is the actual value of point $i$, $\expect{y_{i}}$ is the expected value of that point given by the fit, and $n$ is the total number of points contributing to the fit.

\begin{table}
\begin{center}
\begin{tabular}{cccc}
\hline\hline
& MDR & MLR & SMLR\\
\hline\hline
${\rm a}$ & $0.36\pm0.13$ & $0.52\pm0.17$ & $0.33\pm0.11$\\
${\rm b}$ & $6.24$ & $4.69$ & $6.51$\\
$\sigma$ & 0.32 & 0.31 & 0.31\\
\hline\hline
\end{tabular}
\caption{The Buckley-James regression parameters for the observed compact group sample.}\label{buckleyjames}
\end{center}
\end{table}

When only the detected groups are considered, the least squares fits change only slightly, as is shown in Table~\ref{detectedgroups} (see equation~\ref{fiteq}).
\begin{table}
\begin{center}
\begin{tabular}{cccc}
\hline\hline
& MDR & MLR & SMLR\\
\hline\hline
${\rm a}$ & $0.35\pm0.13$ & $0.49\pm0.17$ & $0.33\pm0.11$\\
${\rm b}$ & $6.55\pm1.26$ & $5.12\pm1.68$ & $6.65\pm1.12$\\
$\sigma$ & 0.32 & 0.31 & 0.31\\
\hline\hline
\end{tabular}
\caption{The least-squares regression parameters for the detected groups in the compact group sample.}\label{detectedgroups}
\end{center}
\end{table}

\subsubsection{Flux Contamination}

Consideration must also be given to the possibility that the compact group observations have measured flux in the area surrounding the compact groups due to bright galaxies not listed in the Hickson catalogue, or within the groups due to dwarf galaxies too faint to be considered members. The former case could occur if the angular extent of the group was less than a third of the beamsize. In this case, the H82 isolation criteria would allow bright interlopers to add \HI flux to the observation. The latter case could occur at any time, as only galaxies within three magnitudes of the brightest galaxy in the group are counted as members. As dwarf galaxies can potentially have significant flux, if they are not correctly accounted for, the content estimation may not be reliable.

\begin{figure}
\begin{center}
\begin{tabular}{cc}
\includegraphics[scale=0.3]{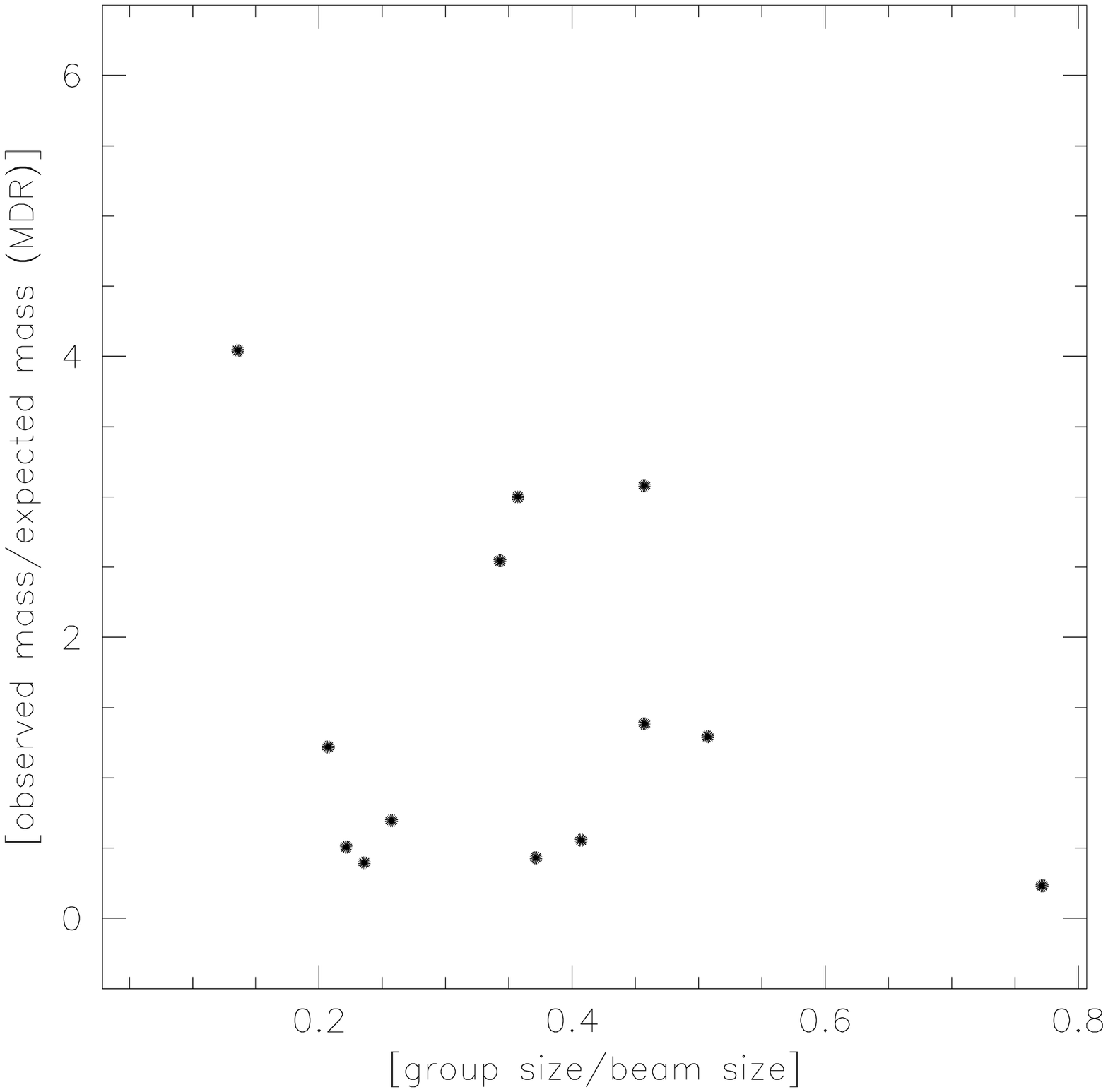} & \includegraphics[scale=0.3]{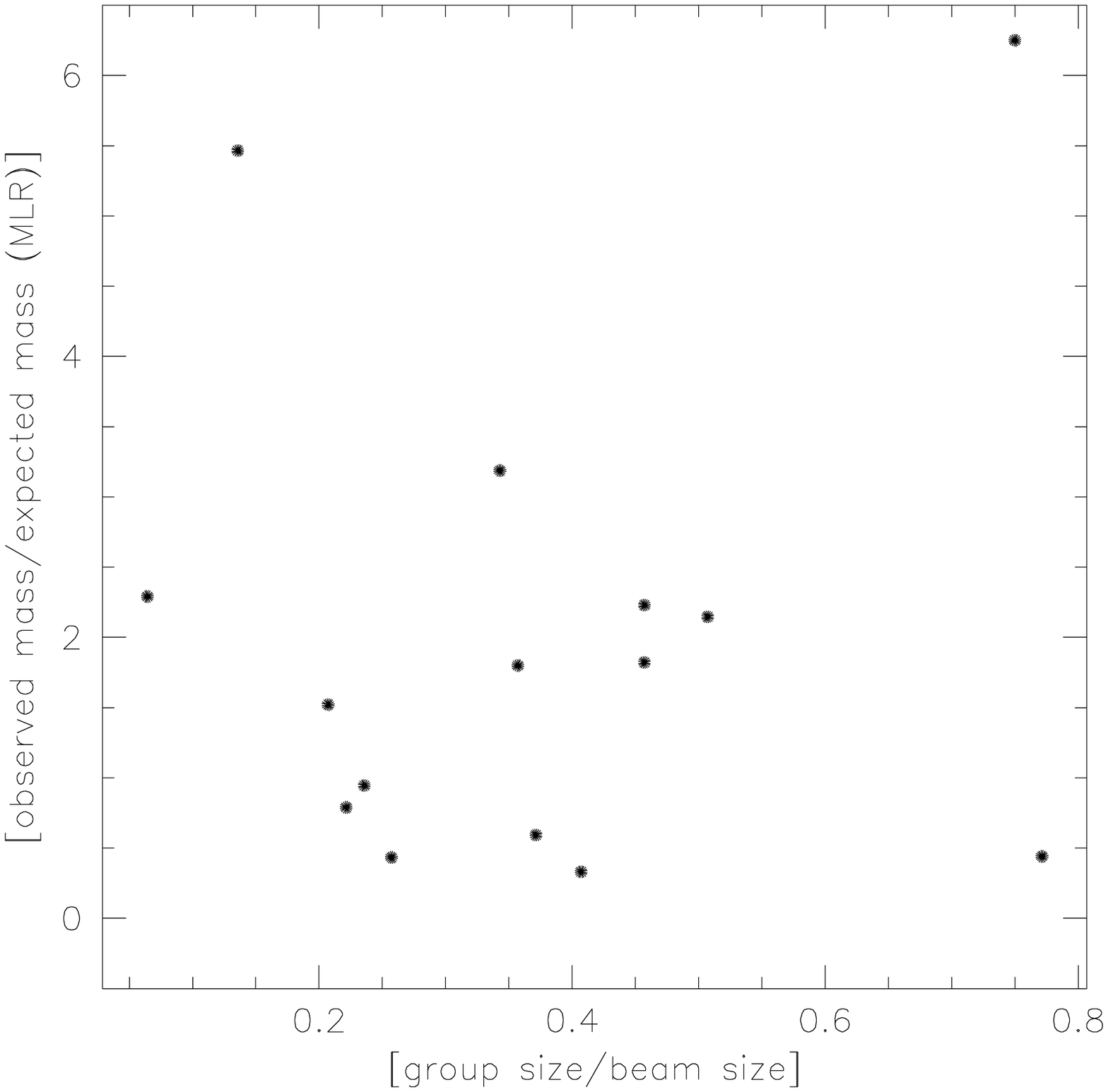}\\
(a) & (b)
\end{tabular}
\caption{The observed-to-expected mass ratio is plotted against the fraction of the beam each compact group fills for (a) expected masses from the MDR, (b) expected masses from the MLR. This measures the effect interlopers outside the compact groups have on the observed masses.}\label{interlopers}
\end{center}
\end{figure}

The effect of any bright interlopers can be estimated by looking at the ratio of the group angular size to the beam size, and whether it correlates with the ratio of the observed flux to the estimated flux. If, at beam-filling ratios below one-third, a larger observed mass is observed with no increase in expected mass, then bright interlopers may be the cause. The plots of beam-filling ratio versus the mass ratio are shown in Figure~\ref{interlopers} for the MDR and the MLR.

The MDR plot of Figure~\ref{interlopers} does not include an outlying group, HCG 26, to keep a reasonable scale for this plot. HCG 26 also appears anomalous in the MLR plot (topmost-right point). This group has many irregular galaxies and a diffuse dominant Scd which makes it difficult to estimate the diameters, and to a lesser extent the luminosities, of the group members. This anomalous group has a large angular extent however, and would not be affected by galaxies outside the group.

For beam-filling ratios of between 0 and 0.3, there is a chance galaxies outside the group may interfere. However, from Figure~\ref{interlopers}, there does not appear to be a significant difference in the mass ratios for the smaller groups when compared to the larger ones. There appear to be two anomalous groups using the MDR estimate, and one with the MLR estimate. The compact group HCG 31 is only anomalous using the MDR mass estimate, and this may be because two galaxies in this group are aligned along the line-of-sight, making the estimation of their diameters difficult, while the sum of their luminosities is not affected. HCG 22 is an outlier using both mass estimators, and has $\sim 5$ times as much mass as expected. This group has several faint galaxies within the group radius which are not designated as group members due to the luminosity criteria. Also a large elliptical galaxy, and faint galaxies lie well outside the group boundaries but within the observing beam. The \HI profile of HCG 22 is quite narrow though, so the velocity dispersion of the group must be low. It is not clear what is causing the underestimation of the mass of the group in this case.

The other four groups which are small enough to be affected by interlopers have the same observed-to-expected mass ratio as the larger groups, suggesting that interloper interference is not especially important. This may be because the beam would lessen the effect the contributions from outlying galaxies could have on the mass, or it may be that interlopers are not present. A search of the HIPASS Catalogue (Meyer et al. 2004) reveals that only two of these groups are catalogued from HIPASS, and both only have a single source within the observing beam.

An estimate of how many low-luminosity galaxies are present in the groups can be made using the compact group luminosity function, such as that found by Zepf et al. (1997). Zepf et al. (1997) surveyed 17 HCGs to look for galaxies which were not included in the H82 group membership, and derived a luminosity function from the results. The luminosity function had a Schechter form, with parameters $M_{*}=-19.5+5\log h$ and $\alpha=-1.0$, where $M_{*}$ is the ``knee'' of the Schechter function, and $\alpha$ is the slope of the function at the faint end.

The amount of mass a number of faint galaxies could contribute to the group can be estimated in the following way. To begin, the brightest galaxy which would not be classified as a group member must be 3 magnitudes fainter than the brightest galaxy in the group, or in terms of luminosity, 15 times fainter. The faint galaxy's contribution to the mass can be computed in the following way:
\begin{eqnarray}
R & \equiv & \frac{\expect{{\rm M}_{\rm HI}({\rm bright})}}{\expect{{\rm M}_{\rm HI}({\rm faint})}} \nonumber \\
 & = & \frac{(1/h^{2})10^{c_{3}(t_{1})}(h^{2}\times15\times {\rm L}_{\rm B})^{c_{4}(t_{1})}}{(1/h^{2})10^{c_{3}(t_{2})}(h^{2}\times{\rm L}_{\rm B})^{c_{4}(t_{2})}} \nonumber \\
 & = & 10^{c_{3}(t_{1})-c_{3}(t_{2})}\frac{(h^{2}\times15\times{\rm L}_{\rm B})^{c_{4}(t_{1})}}{(h^{2}\times{\rm L}_{\rm B})^{c_{4}(t_{2})}} \label{faintmass}
\end{eqnarray}
Here $R$ is the ratio of the predicted \HI mass of the brightest galaxy in the group to the predicted \HI mass of the brightest non-member galaxy, ${\rm L}_{\rm B}$ is the blue luminosity of the brightest non-member galaxy, $t_{1}$ is the morphological type of the brightest member galaxy, and $t_{2}$ is the morphological type of the brightest non-member galaxy.

If we assume average values for $c_{3}$ and $c_{4}$, equation~\ref{faintmass} reduces to $R=(15)^{\overline{c}_{4}}$, where $\overline{c}_{4}$ is an average value of $c_{4}$ and is independent of morphology. From Table~\ref{haynesmlcoeff}, $\overline{c}_{4}=0.67$, thus $R\sim(15)^{0.67}=6.1$. This means that the brightest non-member galaxy would be approximately 6 times less massive than the brightest member galaxy. If the brightest galaxy was not the only member galaxy contributing to the \HI flux, the number of faint galaxies required to cause a significant increase in group mass would be large.

The compact group luminosity function of Zepf et al. (1997) suggests that the number of faint galaxies outnumber the bright galaxies by a factor of between 10 and 100. This provides enough faint galaxies to account for any large \HI excesses. However, Zepf et al. (1997) also found that the faint galaxies have a wider spatial distribution than the bright galaxies. That is, the extent of the group as measured by faint galaxies is larger than the group as defined by H82 by a factor of several.

If it is now assumed that each group, as defined by the extent of the faint non-member galaxies, fills the HIPASS beam entirely, then groups with small angular sizes should again have larger \HI excesses relative to larger groups. This effect is therefore confused with the interloper effect for groups smaller than one-third of the beamsize, but would be the dominant effect in groups larger than this. Since there is no evidence that small groups have a greater excess than large groups (Figure~\ref{interlopers}), faint galaxies do not appear to have a significant effect on the observed mass for the compact group population as a whole. Again, the effect of the beam weighting the contributions from the H82 group more strongly than the surrounding area may be preventing faint galaxies from contributing.

\section{Discussion}\label{discussion}
The purpose of these observations has been to test the null hypothesis that a compact group's \HI mass can be calculated from the sum of the individual galaxies' masses, and that these galaxies have \HI contents matching that of field galaxies with the same observed optical properties.

The Buckley-James fit shown in Table~\ref{buckleyjames} and in Figure~\ref{expectplots} clearly does not agree with the null hypothesis for any of the estimation methods. The slope of the line would suggest that groups with a high expected mass tend to be deficient in \HIn, while those with a low expected mass are not.
However, this result is mostly due to the lack of sensitivity of HIPASS to low mass galaxies.

The least-squares fit parameters to the detected groups are, within errors, the same as the Buckley-James parameters. This is not surprising, since the Buckley-James fits are calculated using the distribution of the known points to estimate the upper limits.

These results show that the groups detected by these observations have \HI contents similar to the reference field sample of HG84, primarily because all the detected groups lie within the area populated by the HG84 reference sample of field galaxies in the expected-observed mass plots. The slope of the line fitted to the detected groups does not match the expected slope, but this may be because the slope is being constrained only at the high expected-high observed mass end (log M$_{\rm HI}>9.5$, log $\expect{{\rm M}_{\rm HI}}>9.5$).

Although the fits from each of the methods lie well within the sample region, because the majority of the groups are undetected by this survey, no definite conclusion about the \HI content of the HCGs can be made. It seems that the limits of HIPASS are not low enough to properly assess the \HI content of the lower mass compact groups.

It is interesting then to determine what the detection limit needs to be to make a more definitive statement about the compact groups as a whole. As was mentioned before, the mass detection limit needs to be lowered to probe the low expected-low observed mass region. To move all the upper limits down to below the equality line would require an increase in sensitivity of 1.5 orders of magnitude, but to move half of them below the equality line only requires an increase in sensitivity of 1 order of magnitude.

If the undetected groups were detected with this sensitivity, then a much tighter constraint could be placed on the slope of the fit, and thus on the \HI content of the groups. If they remained undetected, there would still be enough of a constraint at the low expected-low observed mass end to make an inference about the fraction of the population which could be considered to have normal \HI content.

Practically, the extra sensitivity would require a detection limit of $0.5\sigma$ from HIPASS, which corresponds to 6.5 mJy. If this was the $5\sigma$ level, then the spectral RMS of the observation would need to be 1.3 mJy. To get this noise level would require a 100-fold increase in the integration time, if the same instrumental setup as HIPASS was used.

In the narrowband observations, a spectral RMS of $\sim7$ mJy was obtained for most groups after 14 minutes of integration time. To get a RMS of 1.3 mJy would therefore take approximately 30 times longer, or about 6-7 hours per source. 

The narrowband sample can also be extended by incorporating the results of H97 to provide more confirmed detections. Of the 17 groups not detected here, H97 detects 10, but only the 5 detections with the highest signal-to-noise ratios are used here. The properties of these HCGs are listed in Table~\ref{huchtadd}. When the H97 data is included, the detected groups all still lie within the HG84 sample region. In fact, the H97 detections lie only slightly below the estimated upper limits from HIPASS (see Figure~\ref{comparison}). The resultant slope is slightly steeper though, with $a=0.37$ for the MDR, $a=0.55$ for the MLR, and $a=0.35$ for the SMLR. Only one H97 detection (HCG 42) lies in the low expected-low observed mass region however.

\begin{table*}
\begin{center}
{\tiny
\begin{tabular}{ccccccc}
\hline\hline
Name & RMS & ${\rm S}_{\rm int}$ & $v_{\rm LSR}$ & $v_{50}$ & $v_{20}$ & ${\rm M}_{\rm HI}$\\
 & (mJy) & (Jy km s$^{-1}$) & (km s$^{-1}$) & (km s$^{-1}$) & (km s$^{-1}$) & (10$^{9}$ M$_{\odot}$)\\
\hline\hline
HCG 003 & 2.3 & 6.2$\pm$1.9 & 7986$\pm$47 & 403$\pm$93 & 1057$\pm$146 & 18.9$\pm$5.9\\
HCG 014 & 1.1 & 3.2$\pm$0.4 & 5476$\pm$2 & 1217$\pm$3 & 1231$\pm$5 & 4.6$\pm$0.6\\
HCG 042 & 4.6 & 3.7$\pm$1.3 & 3994$\pm$18 & 201$\pm$35 & 307$\pm$55 & 2.8$\pm$1.0\\
HCG 043 & 2.0 & 3.1$\pm$0.7 & 9973$\pm$8 & 352$\pm$17 & 426$\pm$26 & 14.9$\pm$3.4\\
HCG 091 & 2.8 & 9.7$\pm$1.9 & 7242$\pm$9 & 483$\pm$18 & 634$\pm$29 & 24.4$\pm$4.8\\
\hline\hline
\end{tabular}}
\caption{The \HI properties of the HCGs not detected by HIPASS, but detected by H97.}\label{huchtadd}
\end{center}
\end{table*}

Finally, this result can be contrasted to that found by Solanes et al. (2001) who studied the \HI deficiency in spiral galaxies found in clusters. They found that within 1 Abell radius (1 R$_{\rm A}$) of the cluster centre, spiral galaxies show strong deficiency, while in the outer parts of the cluster, the \HI contents were similar to those found in the field. As the galactic densities of compact groups can approach that of the cores of rich clusters (H82), it is interesting then that compact groups do not show the same deficiency.

Solanes et al. (2001) concluded that hydrodynamical effects caused by the interaction of the \HI with the hot intracluster medium in the centre of the clusters was the cause of the \HI deficiency. They also found \HI stripping was a relatively recent event (a few Gyr ago) for the deficient galaxies, as the optical properties of the galaxies have not been affected.

One possibility then for the normal \HI content of the compact group sample investigated in this paper is the lack of a hot intragroup medium to strip the \HIn. This agrees with X-ray observations of the HCGs from Ponman et al. (1996), who concluded that only 20\% of groups had diffuse X-ray emission with luminosity greater than 10$^{42}$ ergs. This luminosity is an order of magnitude weaker than the weakest cluster emission reported by Solanes et al. (2001).

\section{Conclusions}

\begin{itemize}
\item{This study has shown that southern compact groups in the HCG catalogue appear have a \HI content consistent with the hypothesis that they are comprised of galaxies similar to those found in the field.}
\item{In comparing the mass estimation methods of HG84 and H97, it was found that they all give similar results, although the scatter in the relationships prevents them from predicting the mass of an isolated galaxy accurately.}
\item{The observations of the detected compact groups all have measured \HI contents within the predicted range for their expected content. However, the majority of groups were not detected in this study, and so the \HI content of the compact group population as a whole is not yet known.}
\item{The survey has not been significantly affected by biasing due to interlopers lying far outside the group radius; nor from faint galaxies not listed as part of the compact groups, but contributing to the \HI mass.}
\item{The \HI in the compact groups may have evolved through interactions, and star-formation, but it is clear that the fraction of \HI consumed is not large. It is possible that \HI depletion is correlated with morphological evolution towards earlier types, which would reduce the expected \HI mass of the group.}
\item{A survey of these groups with much higher sensitivity, leading to more detections or more stringent upper limits, will give a clearer picture on their \HI content.}
\end{itemize}

\section{Acknowledgments}
The authors would like to thank the HIPASS team for the data used for our detection experiment, and the staff at the Parkes radio telescope for assistance/troubleshooting during our observing runs. Thanks also to Donald Payne, Martin Zwaan, and Stacey Mader for observing assistance. Thanks to Martin Meyer for the use of {\bf mmspect}, and for providing us with the HIPASS catalogue. JS thanks Melbourne University for financial assistance through a Melbourne University Postgraduate Scholarship. The Parkes telescope is part of the Australia Telescope, which is funded by the Commonwealth of Australia for operation as a National Facility managed by CSIRO. This work has made use of the MIRIAD software package. This research has made use of the NASA/IPAC Extragalactic Database (NED) which is operated by the Jet Propulsion Laboratory, California Institute of Technology, under contract with the National Aeronautics and Space Administration.

\section*{References}

\reference Barnes, D. G., 1998, in ASP Conf. Ser., Vol. 145, Astronomical Data Analysis Software and Systems VII, ed. R. Albrecht, R. N. Hook, H. A. Bushouse (San Francisco: Astronomical Society of the Pacific), p 32
\reference Barnes, D. G., Staveley-Smith, L., de Blok, W. J. G., Oosterloo, T., Stewart, I. M., Wright, A. E., Banks, G. D., Bhathal, R. et al., 2001, MNRAS, 322, 486
\reference Braun, R., \and Burton, W. B., 1999, A\&A, 341, 437
\reference Buckley, J., \and James, I., 1979, Biometrika, 66, 429
\reference Chamaraux, P., 1987, A\&A, 177, 326
\reference Chamaraux, P., Balkowski, C., \and Fontanelli, P., 1986, A\&A, 165, 15
\reference de Vaucouleurs, G., de Vaucouleurs, A., \and Corwin, H. G., 1976, in Second Reference Catalogue of Bright Galaxies, (Austin: University of Texas Press)
\reference Haynes, M. P., \and Giovanelli, R., 1984, AJ, 89, 758
\reference Hibbard, J. E., \and Van Gorkom, J. H., 1996, AJ, 111, 655
\reference Hickson, P., 1982, ApJ, 255, 382
\reference Hickson, P., 1993, in Atlas of Compact Groups of Galaxies, (New York: Gordon and Breach Science Publishers)
\reference Hickson, P., 1997, ARA\&A, 35, 357
\reference Huchtmeier, W. K., 1997, A\&A, 325, 473
\reference Huchtmeier, W. K., \and Richter, O. G., 1988, A\&A, 203, 237
\reference Isobe, T., Feigelson, E. D., \and Nelson, P. I., 1986, ApJ, 306, 490
\reference Karachentseva, V. E., 1973, Astrofizicheskie Issledovaniia Izvestiya Spetsial'noj Astrofizicheskoj Observatorii, 8, 3
\reference Leon, S., Combes, F., \and Menon, T. K. 1998, A\&A, 330, 37
\reference Meyer, M., Zwaan, M., Webster, R.L., Staveley-Smith, L., Ryan-Weber, E., Drinkwater, M.J., Barnes, D.G., Howlett, M. et al. 2004, MNRAS, 350, 1195
\reference Nilson, P., 1973, in Uppsala General Catalogue of Galaxies, (Acta Universitatis Upsaliensis. Nova Acta Regiae Societatis Scientiarum Upsaliensis - Uppsala Astronomiska Observatoriums Annaler, Uppsala: Astronomiska Observatorium)
\reference Ponman, T., Bourner, P. D. J., \and Ebeling, H., 1996, Roentgenstrahlung from the Universe, 357
\reference Prandoni, I., Iovino, A., \and MacGillivray, H. T., 1994, AJ, 107, 1235
\reference Sadler, E. M., Oosterloo, T., \and Morganti, R., 2002, in The Dynamics, Structure \& History of Galaxies, (San Francisco: Astronomical Society of the Pacific), 215
\reference Sandage, A., \and Tammann, G. A., 1981, in Revised Shapley-Ames Catalog of Bright Galaxies, (Washington, D.C.: Carnegie Institute of Washington)
\reference Schneider, S. E., Helou, G., Salpeter, E. E., \and Terzian, Y., 1986, AJ, 92, 742
\reference Solanes, J. M., Manrique, A., Garcia-Gomez, C., Gonzalez-Casado, G., Giovanelli, R., \and Haynes, M. P., 2001, ApJ, 548, 97
\reference Verdes-Montenegro, L., Yun, M. S., Perea, J., del Olmo, A., \and Ho, P. T. P. 1998, ApJ, 497, 89
\reference Verdes-Montenegro, L., Yun, M. S., Williams, B. A., Huchtmeier, W. K., Del, Olmo A., \and Perea, J., 2001, A\&A, 377, 812
\reference Williams, B. A., \and Rood, H. J., 1987, ApJS, 63, 265
\reference Zepf, S. E., de Carvalho, R. R., \and Ribeiro, A. L. B. 1997, ApJL, 488, 11

\end{document}